\documentclass[12pt, notitlepage,aps,preprintnumbers,superscriptaddress,nofootinbib,aps,prd,floatfix]{revtex4-2}
\pdfoutput=1
\usepackage{graphicx,diagbox}
\usepackage{subfigure}
\usepackage{amsthm} 
\usepackage{amsmath,amssymb,physics}
\usepackage{url} 
\usepackage{slashed}
\usepackage{xcolor} 
\usepackage{multirow}
\usepackage{hyperref}
\usepackage{soul}
\definecolor{nicered}{rgb}{0.5,0.,0.}
\definecolor{nicegreen}{rgb}{0.,0.5,0.}
\definecolor{niceblue}{rgb}{0.,0.,0.5}
\hypersetup{colorlinks,citecolor=nicegreen,linkcolor=nicered,urlcolor=niceblue}
\newcolumntype{C}[1]{>{\centering\let\newline\\\arraybackslash\hspace{0pt}}m{#1}}
\usepackage{orcidlink}
\usepackage{enumitem} 
\usepackage[normalem]{ulem}
\setlist{nolistsep} 
\usepackage{setspace}
\newcommand{\GeV}{\textrm{GeV}}
\newcommand{\TeV}{\textrm{TeV}}

\begin{document}
\preprint{PITT-PACC-2407, MSUHEP-24-016}
\title{Colorful Particle Production at High-Energy Muon Colliders}

\author{Tao Han\,\orcidlink{0000-0002-5543-0716}}
\affiliation{Pittsburgh Particle Physics Astrophysics and Cosmology Center,
Department of Physics and Astronomy,
University of Pittsburgh, Pittsburgh, PA 15260, USA\looseness=-1}

\author{Matthew Low\,\orcidlink{0000-0002-2809-477X}}
\affiliation{Pittsburgh Particle Physics Astrophysics and Cosmology Center,
Department of Physics and Astronomy,
University of Pittsburgh, Pittsburgh, PA 15260, USA\looseness=-1}

\author{Tong Arthur Wu\,\orcidlink{0009-0002-9485-3938}}
\affiliation{Pittsburgh Particle Physics Astrophysics and Cosmology Center,
Department of Physics and Astronomy,
University of Pittsburgh, Pittsburgh, PA 15260, USA\looseness=-1}

\author{Keping Xie\,\orcidlink{0000-0003-4261-3393}}
\affiliation{Pittsburgh Particle Physics Astrophysics and Cosmology Center,
Department of Physics and Astronomy,
University of Pittsburgh, Pittsburgh, PA 15260, USA\looseness=-1}
\affiliation{Department of Physics and Astronomy, Michigan State University, East Lansing, MI 48824, USA\looseness=-1
\vspace{1em}}

\begin{abstract}
A high-energy $\mu^+\mu^-$ collider provides a wide variety of mechanisms for the production of new heavy particles. While the reach for such particles via the direct annihilation of $\mu^+\mu^-$ will approach the center-of-mass energy of the collider, the partonic fusions from gauge bosons, quarks, and gluons, originating from the incoming muon beams will open new channels for single production and pair production of particles with different quantum numbers.  We present the production rates for a wide variety of colored states including color-triplets, color-sextets, color-octets, leptoquarks, and leptogluons.  We find that pair production from the direct annihilation of $\mu^+\mu^-$ generally has a cross section of $0.1 - 1$ fb once above the production threshold.  On the other hand, pair production through the quark and gluon content of the muon leads to a cross section of roughly $10^{-4}$ fb at the same particle mass.  We perform simple estimations of the mass reach for each particle and find that a 10 TeV muon collider can extend the reach for color-triplets beyond what is possible at the high luminosity run of the Large Hadron Collider.  Leptoquarks and leptogluons, with sensitivity driven by single production, can also be probed to higher masses at a muon collider than what the Large Hadron Collider can reach.  A final example where a muon collider has superior reach is for color-octet scalars and vectors.  Together, these cases illustrate the point that a muon collider is a competitive machine for searching for colored heavy particles, thus strengthening the motivation for such lepton colliders in the energy frontier. Although our study is focused on a muon collider, our results are largely applicable to high-energy $e^+e^-$ collisions as well.
\end{abstract}

\maketitle
\tableofcontents

\section{Introduction}
\label{sec:intro}

Building upon the highly successful program at the CERN Large Hadron Collider (LHC), signified by the Higgs boson discovery~\cite{ATLAS:2012yve,CMS:2012qbp} and the extensive searches for new particles and interactions beyond the Standard Model (SM), the next target in the high-energy frontier is to explore physics with a partonic center-of-mass energy around 10 TeV~\cite{P5:2023wyd}.  The prospect of constructing a multi-TeV $\mu^+ \mu^-$ collider~\cite{Delahaye:2019omf,Bartosik:2020xwr,Schulte:2021hgo,Long:2020wfp} has garnered significant attention within the particle physics community~\cite{MuonCollider:2022xlm,Aime:2022flm,Black:2022cth,Accettura:2023ked}.  With the collisions of two elementary particles at high energies, a multi-TeV muon collider would open a new energy threshold for heavy particle production via $\mu^+ \mu^-$ direct annihilation by fully utilizing the collider center-of-mass energy. Additionally, the collinear radiation of electroweak (EW) gauge bosons off the colliding beams~\cite{Ciafaloni:2000df,Ciafaloni:2001mu,Chen:2016wkt} enhances the production through the vector-boson fusion mechanism~\cite{Costantini:2020stv,BuarqueFranzosi:2021wrv}. In this sense, a muon collider effectively becomes a vector boson collider~\cite{Costantini:2020stv,BuarqueFranzosi:2021wrv,Han:2020uid,AlAli:2021let} and, in fact, a collider of any EW parton since they are all dynamically generated~\cite{Bauer:2017isx,Fornal:2018znf,Bauer:2018arx,Han:2020uid,Han:2021kes,Garosi:2023bvq}.  As a consequence, a large variety of channels of different spins and electroweak and color quantum numbers can emerge in both initial and final states. 

New particles charged under the color of Quantum Chromodynamics (QCD) would be readily produced at the LHC and future hadron colliders via the scattering of quarks ($q$) and gluons ($g$) once the mass threshold of the new particles is crossed.  The current bounds of LHC Run 2 on the masses of gluino and light-flavor squarks in supersymmetric theories are around $2.2 - 2.4~\TeV$ \cite{ATLAS:2020syg,ATLAS:2023afl} and $1.3 - 1.9~\TeV$ \cite{ATLAS:2020syg}, respectively, while the bottom-flavor and top-flavor squark limits only reach around $0.5 - 1.3~\TeV$ \cite{CMS:2023ktc,ATLAS:2024rcx}, depending on the assumptions and decay channels used in specific analyses~\cite{ParticleDataGroup:2024cfk}. The mass bound on a vector-like quark ranges from $0.7 - 2~\TeV$, depending on the corresponding decay width~\cite{Banerjee:2024zvg}.  The high luminosity LHC (HL-LHC) may extend the mass reach for many particles, like raising the top squarks sensitivity to 2 TeV~\cite{ATLAS:2018zrp,ATLAS:2019mfr,Baer:2023uwo}, however, due to the large Standard Model QCD backgrounds, those bounds depend on the assumption that the new heavy particles decay to observable final states, such as missing transverse energy, and energetic leptons and jets.  Not to mention that these final decay products need to be sufficiently energetic to be distinct from the background processes in the SM.

A high-energy muon collider, on the other hand, may provide complementary information because of the clean experimental environment for signal reconstruction, the variety of the available channels to search for, and the well-constrained kinematics for the possible determination of the particle properties such as the mass and spins, as we will demonstrate. 
Some work has been devoted to searching for heavy-colored particles in muon colliders in the literature~\cite{Berger:1996un,AlAli:2021let,Asadi:2021gah,Huang:2021biu,Bandyopadhyay:2021pld,Qian:2021ihf,Lv:2022pts,Azatov:2022itm,Belyaev:2023yym,Ghosh:2023xbj,Liu:2023jta,Bhaskar:2024snl}.
In this work, we present the production cross sections of colored states and discuss their discovery potential at a muon collider.   We cover the most commonly occurring colored particles in BSM theories including color-triplets, color-sextets, leptoquarks, leptogluons, and color-octets.  We find that a 10 TeV muon collider has superior sensitivity, compared to the HL-LHC, to color-triplet fermions and scalars, scalar leptoquarks, fermionic leptogluons, and color-octet scalars and vectors.  In cases where the limits are driven by pair production, the mass reach approaches the kinematic limit of 5 TeV, whereas in cases where the limits are driven by single production, the mass reach is coupling-dependent, but typically reaches upward of 8 TeV for large enough couplings.
We would like to emphasize that although our studies are fully focused on a muon collider, our results are largely applicable to high-energy $e^+e^-$ collisions as well.
 
The rest of the paper is organized as follows.  In Sec.~\ref{sec:heavy}, we first present several representative color states that are well motivated in theories beyond the Standard Model.  We then describe the general formalism for heavy particle production.  In Sec.~\ref{sec:EWprod}, we explore heavy particle production through EW scattering, which is followed by leptonquark/leptogluon production in Sec.~\ref{sec:fusion}, and by QCD scattering in Sec.~\ref{sec:QCD}.  Conclusions and future prospects are given in Sec.~\ref{sec:conclusion}.

\section{Heavy Colored Particle Production in $\mu^+ \mu^-$ Collisions}
\label{sec:heavy}

\subsection{Colored States}

Theories beyond the SM often predict the existence of heavy-particle states charged under QCD color.  Weak-scale Supersymmetry (SUSY)~\cite{Nilles:1983ge,Haber:1984rc,Martin:1997ns} is among the best motivated theories for physics beyond the SM.  SUSY predicts partners for each of the SM states which implies that there is a color-octet fermion known as the ``gluino'' ($\tilde{g}$), and color-triplet scalars known as ``squarks'' ($\tilde{q}$).  Naturalness considerations prefer that the top-flavor squark, the ``stop'' ($\tilde{t}$), and the gluino are not much heavier than the EW scale, at the order of a few TeV~\cite{Buchmueller:2013exa,Papucci:2011wy}.  Thus, a multi-TeV muon collider would cover the mass range of interest and could even potentially study the SUSY partner properties. In addition to the electroweak production of squark pairs, squarks and gluinos can be produced via QCD from the partonic content of the quark and gluon of the muon in a high-energy muon collider~\cite{Han:2021kes}. 

In contrast to supersymmetry, composite Higgs theories often contain a color triplet fermion ($T$)~\cite{Agashe:2004rs,Schmaltz:2005ky,Panico:2015jxa}, associated with the top quark. It typically has vector-like interactions with the SM and couples to the electroweak sector via mixing. The leading production mechanism for vector-like fermionic states is pair production in the $s$-channel through $\gamma$ and $Z$.  Color-triplet vectors ($\omega$) appear as $\rho$-like resonances in theories of vector-like confinement~\cite{Kilic:2009mi,Arkani-Hamed:2016kpz}. 

Moving beyond color triplets, grand unified theories often lead to colored states with exotic quantum numbers, such as color-triplet ``leptoquarks'' ($\ell_q$), color-sextet ``diquarks,'' and color-octet ``leptogluons'' ($\ell_g$) in superstring-inspired $E_6$ models~\cite{Hewett:1988xc}.  Other exotic colored states exist in theories as scalar octets ($S_8$) in supersymmetric theories~\cite{Plehn:2008ae} or as heavy pions in theories of vector-like confinement~\cite{Kilic:2009mi,Arkani-Hamed:2016kpz}.  Another interesting state is the vector octet ``axigluon'' $(V_8)$ in $SU(3) \times SU(3)$ theories~\cite{Frampton:2009rk} or vector-octet $\rho$-like resonance in theories of vector-like confinement~\cite{Kilic:2009mi,Arkani-Hamed:2016kpz}.

While we borrow our naming convention from the most common beyond the SM incarnations, the calculated rates are general and do not depend on model details (except in special cases which are explicitly specified).  We compile a list in Table \ref{tab:Model2} for representative colored particle states that we study. We have also included the current lower bounds on the masses from existing direct searches. 

\begin{table} [tb]
  \centering
  \begin{tabular}{c|c|c|c|c|c}
    \hline
     Model & Label  & Spin & $SU(3)_c$  & $|Q_e|$  & Current Limit (TeV) \\
     \hline
    Squark &  $\tilde{t} \; (\tilde{b})$ & 0 &  3  & 2/3 (1/3) &  1.3~\cite{ATLAS:2024rcx} \\
    Scalar leptoquark &  $S_{\ell q}$  & 0  & 3   &  ${}^\dagger$ & 1.4-2.0~\cite{Bhaskar:2023ftn,CMS:2024bnj}  \\
    Scalar octet / Technipion & $S_8$ & 0 & 8 & 0,1 & 1.25~\cite{Darme:2021gtt,Miralles:2019uzg} / 0.35~\cite{ATLAS:2020zzb} \\
    Scalar diquark & $D_6$ & 0 & 6 & 2/3, 4/3 &  7.5~\cite{CMS:2019gwf}  \\
    \hline
    Vector-like quark & $T \; (B)$ & 1/2 & 3 & 2/3 (1/3) & 2.1~\cite{ATLAS:2024xdc}(1.59~\cite{ATLAS:2022tla}) \\
     Gluino & $\tilde{g}$ & 1/2 & 8 & 0 & 2.2~\cite{ATLAS:2023afl} \\
     Leptogluon & $\ell_{g}$ & 1/2 & 8 & 1 &  2.5~\cite{Mandal:2012rx,Mandal:2016csb}\\
     Excited quark & $Q_6$ & 1/2, 3/2 & 6 &1/3, 2/3 &  6.3~\cite{CMS:2019gwf}  \\
\hline
   Vector leptoquark &  $V_{\ell q}$ & 1 & 3   & $1/3-5/3$ &  2.12~\cite{CMS:2024bnj} \\
   Vector octet / Techni-$\rho$ & $V_8$ &1 & 8 &0,1 &  1.6~\cite{Darme:2021gtt} / 2~\cite{ATLAS:2020zzb} \\
   Vector diquark & $V_6$ & 1   & 6 &2/3, 4/3 &  3.42~\cite{Das:2015lna} \\
    \hline
    \end{tabular}
    \caption{Summary for colored particles in beyond the SM scenarios and the corresponding lower mass bounds [TeV]. $^\dagger$The charges of the scalar leptoquarks can be found in Table~\ref{tab:SLQ}.}
    \label{tab:Model2}
\end{table}

\subsection{Heavy Particle Production Formalism}
\label{sec:states}

\begin{figure}
  \centering
  \includegraphics[width=0.35\linewidth]{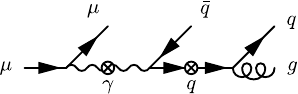}
  \caption{Representative collinear splittings for $\mu\to\mu\gamma$, and $\gamma\to q\bar{q}$, and $q\to qg$, shown sequentially.}
  \label{fig:split}
\end{figure}

Similar to QCD parton distribution functions (PDFs) in hadronic collisions at the LHC, electroweak PDFs, along with the corresponding factorization formalism, are introduced to describe the partonic collisions at high-energy muon colliders~\cite{Han:2020uid,Han:2021kes,Garosi:2023bvq}.  While QCD PDFs are inherently non-perturbative at low energies, EW PDFs are perturbatively calculable, thanks to the perturbative nature of the EW theory. Even though muons do not couple directly to colored particles, through higher-order splittings as illustrated in Fig.~\ref{fig:split}, quarks and gluons can be dynamically generated as components of the EW PDFs~\cite{Han:2021kes}.  As such, a high-energy muon collider can provide a very rich environment with a large variety of initial states for new particle production. 

\begin{figure}
  \subfigure[]{\includegraphics[width=0.23\textwidth]{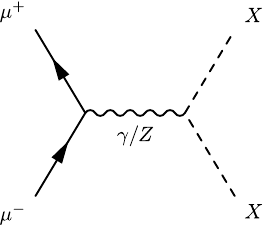}\label{fig:DY}}
  \qquad\qquad
  \subfigure[]{\includegraphics[width=0.26\textwidth]{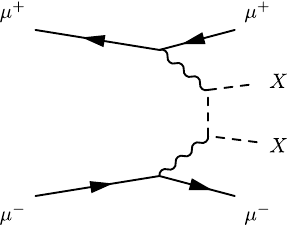}\label{fig:VBF}}
  \qquad\qquad
  \subfigure[]{\includegraphics[width=0.27\textwidth]{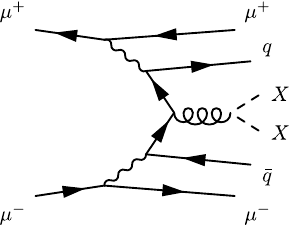}\label{fig:qqbQCD}}
  \caption{Representative Feynman diagrams for the pair production of a new charged and colored particle $X$ (represented with a dashed line) through (a) $\mu^+\mu^-$  annihilation, (b) vector-boson fusion, and (c) quark/gluon scattering.  Here, the initial quarks and gluons come from the high-order splitting, as illustrated in Fig.~\ref{fig:split}.}
  \label{fig:feynman_production_EW}
\end{figure}

Under the assumption of electroweak factorization, the semi-inclusive cross section $\sigma(\mu^+\mu^- \to X X)$ for the pair production of the colored particle $X$ at a high-energy muon collider can be formulated as the convolution of the partonic cross section $\hat{\sigma}(ij\rightarrow X X)$ and the parton luminosity $\dd\mathcal{L}_{ij}/\dd\tau$\footnote{Our notation of $XX$ should be understood to be $XX$ when $X$ is its own antiparticle and $X\bar{X}$ when $\bar{X}$ is the antiparticle of $X$.}
\begin{equation}
\sigma(\mu^+\mu^- \to X X)=
 \int_{\tau_{0}}^{1} \dd\tau  \sum_{ij}\frac{\dd\mathcal{L}_{ij}}{\dd\tau}\  \hat{\sigma}(ij\rightarrow X X).
\end{equation}
Here $ij$ are the labels of the incoming partons.  The parton luminosity is independent of the partonic process and is defined as
\begin{equation}
\frac{\dd\mathcal{L}_{ij}}{\dd\tau} = \frac{1}{1+\delta_{ij}}  \int^{1}_{\tau} \frac{\dd\xi}{\xi} 
 \left[ f_{i}(\xi, Q^{2})f_{j}\Big(\frac{\tau}{\xi},Q^{2} \Big) + (i \leftrightarrow j) \right].
\end{equation}
Here $Q^2$ is the factorization scale, $\tau = \hat s/s$ with $\sqrt{s}\ (\sqrt{\hat{s}})$ the center-of-mass energy of the initial beam (partonic) collisions, and the production threshold is at $\tau_{0} = 4m_{X}^{2}/s$.  Some representative Feynman diagrams for $XX$ production are depicted in Fig.~\ref{fig:feynman_production_EW}. 

At leading order (LO), the PDFs of a high-energy lepton can be approximated by the collinear splitting functions of beam particles into the corresponding gauge bosons. More specifically, the photon PDF can be obtained through the $\ell\to\ell\gamma$ splitting $P_{\gamma\ell}$ according to
\begin{equation}
f_{\gamma/\ell}(x,Q^2)=\frac{\alpha}{2\pi}P_{\gamma\ell}(x)\ln\frac{Q^2}{m_\mu^2},
\qquad\qquad
P_{\gamma\ell}(x)=\frac{1+(1-x)^2}{x},
\end{equation}
where $x$ is the momentum fraction carried by the photon.  If the factorization scale $Q$ is taken to be the lepton beam energy $E_\ell$, the LO photon PDF coincides with the equivalent photon  approximation~\cite{vonWeizsacker:1934nji,Williams:1934ad,Budnev:1975poe}.

A similar treatment has been extended to the electroweak gauge bosons $W/Z$ as well, resulting in the effective $W$ approximation~\cite{Kane:1984bb,Dawson:1984gx,Chanowitz:1985hj}.  At one order higher, as illustrated in Fig.~\ref{fig:split}, the gauge bosons can further split into quarks, $\gamma/Z/W\to q\bar{q}$, which provides the quark content of a lepton~\cite{Han:2021kes}.  A gluon parton can subsequently be generated through $q\to qg$.  In such a way, all the SM particles can be dynamically generated through perturbative splittings, leading to much richer production mechanisms for new particles, including vector boson fusion, gluon fusion, and quark annihilation. 

\setlength{\unitlength}{0.6cm}

At high energies, well above the lepton and gauge boson masses, large logarithms are present, $\ln(Q^2/m_\ell^2)$ and $\ln(Q^2/M_{Z,W}^2)$, as a consequence of collinear splittings, which need to be resummed to all orders through the Dokshitzer–Gribov–Lipatov–Altarelli–Parisi (DGLAP) evolution equations~\cite{Gribov:1972ri,Lipatov:1974qm,Altarelli:1977zs,Dokshitzer:1977sg}. Recently, the complete SM DGLAP evolution has been achieved for proton beams~\cite{Bauer:2017isx,Fornal:2018znf,Bauer:2018arx} as well as for lepton beams~\cite{Han:2020uid,Han:2021kes,Garosi:2023bvq}. Based on the quark and gluon PDFs of lepton beams, it was found that a large fraction of dijet production in the SM at lepton colliders emerges from quark and gluon scattering~\cite{Han:2021kes}.  In this work, we present the production of the BSM colored particles at a high-energy muon collider including the quark and gluon initiated production processes.

The main calculations and simulations in this work are performed with the multi-purpose event generators \texttt{MadGraph}~\cite{Alwall:2014hca,Frederix:2018nkq} and \texttt{Whizard}~\cite{Kilian:2007gr,Moretti:2001zz,Christensen:2010wz}, interfaced with the EW PDFs of a muon beam~\cite{Han:2020uid,Han:2021kes} through the \texttt{LHAPDF} convention~\cite{Buckley:2014ana}.  New particles are implemented via {\tt FeynRules}~\cite{Alloul:2013bka} and \texttt{NLOCT}~\cite{Degrande:2014vpa} into the {\tt UFO} format~\cite{Degrande:2011ua}. \texttt{FeynCalc}~\cite{Shtabovenko:2020gxv,Shtabovenko:2016sxi,Mertig:1990an} and \texttt{FeynArts}~\cite{Hahn:2000kx} are also used for cross checking some cross sections.

\section{Electroweak Production}
\label{sec:EWprod}

The major advantage for new heavy particle discovery at a muon collider is that once the center-of-mass energy is above the pair production threshold any new particle that has electroweak couplings will be readily produced in pairs.  For example, the case of color-triplet fermions is shown in Fig.~\ref{fig:DY}.  The first cases we consider are where the colored particles are pair-produced directly from $\mu^+ \mu^-$ initial state annihilation. This occurs when the new colored particles carry non-color gauge charges and couple to a common mediator.  The leading production is mediated by an $s$-channel gauge boson and shown in Fig.~\ref{fig:DY}. Diagrams of this form have a cross section that scales as $\alpha^2$ at a scale of $M_X$.

For these states, the subleading production, shown in Fig.~\ref{fig:VBF}, also contributes to the rate.  Formally, at fixed order, the rate for this diagram scales as $\alpha^4$. However, large logarithms $\alpha \ln^2(Q^2/m_\ell^2)$ from the collinear splittings need to be resummed and lead to the PDFs, as discussed in Sec.~\ref{sec:heavy}.  This production channel increases with the physics scale and may take over at sufficiently high energies. 
We refer to this as $\gamma\gamma$ fusion or vector boson fusion (VBF) more generally.

Finally, there are contributions from QCD production, shown in Fig.~\ref{fig:qqbQCD}.  These contributions have cross sections that scale formally starting at order $\alpha^4 \alpha_s^2$ and are subleading with respect to electroweak production.  On the other hand, for heavy states that do not carry electroweak quantum numbers, these production mechanisms may constitute the leading contribution for color particle production at lepton colliders.  Once again, one needs to properly treat the quarks and gluons in the PDF framework.

In the following subsections we calculate rates for one choice of electroweak quantum numbers, but one can rescale the inclusive cross section for other quantum numbers.  Let the electric charge of the reference choice be $Q_{\rm ref}$, then for pair production, to scale the rate for a particle with electric charge $Q$ the rate is
\begin{equation}\label{eq:power}
\sigma = \left(\frac{Q}{Q_{\rm ref}}\right)^2\sigma_{\mu^+ \mu^-}
+ \left(\frac{Q}{Q_{\rm ref}}\right)^4 \sigma_{\gamma\gamma}
+\sigma_{\rm QCD},
\end{equation}
where $\sigma_{\mu^+ \mu^-}$ is the $\alpha^2$ contribution, $\sigma_{\gamma\gamma}$ is the $\alpha^4$ contribution, and $\sigma_{\rm QCD}$ is the $\alpha^4 \alpha_s^2$ contribution.  Each contribution involves different partonic initial states which should be calculated from the partonic cross sections convolved with the PDFs with the SM DGLAP evolution.

\subsection{Color-Triplet Fermions: Vector-like Quarks}

Perhaps one of the most studied BSM scenarios is a new color-triplet fermion ($T$).  Such a state is common in composite Higgs models~\cite{Panico:2015jxa,Liu:2023jta} and little Higgs models~\cite{Han:2003wu}.  Due to the non-decoupling nature, additional heavy chiral fermions imply a sizeable contribution to the loop-induced Higgs coupling to gluons and photons, which are excluded by Higgs measurements at the LHC, for example through the production rates~\cite{ATLAS:2019nkf,CMS:2018gwt}. As a consequence, such a color-triplet fermion $T$ must be vector-like~\cite{Buchkremer:2013bha,Cacciapaglia:2012dd}, which means they acquire a mass independent of the Higgs field and may or may not receive mass contributions from the Higgs field.  While many choices of electroweak and hypercharge quantum numbers are possible and motivated by particular theories, we choose the case for the gauge quantum numbers (SU(3)$_c \otimes$SU(2)$_L$)$_Y$ as 
\begin{equation*}
\text{color-triplet fermion }(T): 
\quad
(3, 1)_{2/3},
\end{equation*}
which are the same quantum numbers as the right-handed top quark.

\begin{figure}
  \centering
  \includegraphics[width=0.45\textwidth]{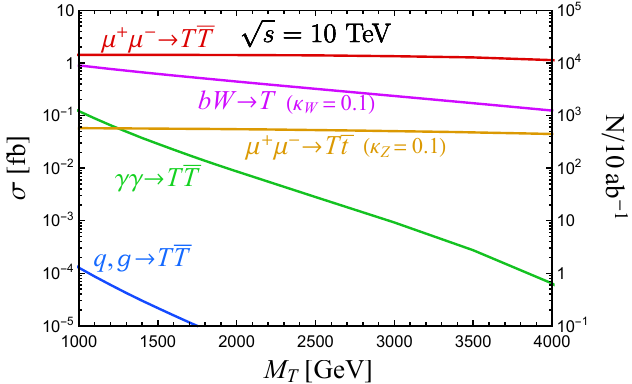}
  \qquad
  \includegraphics[width=0.45\textwidth]{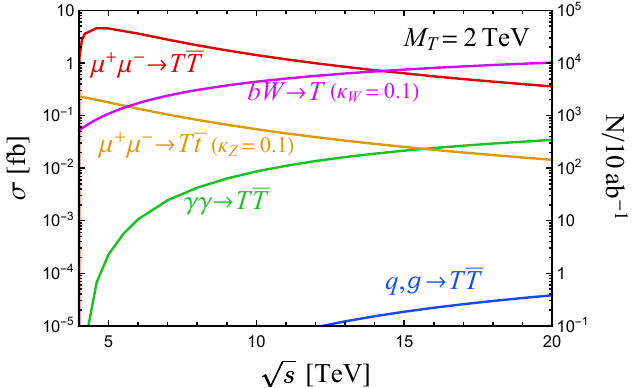}
  \caption{The cross section for the pair production $T\bar{T}$ of a color-triplet fermion $T$ with quantum numbers $(3, 1)_{2/3}$ as a function of mass (left) and as a function of center-of-mass energy (right). The model-independent signals of $\mu^+ \mu^-$ annihilation (red), $\gamma\gamma$ fusion (green), and QCD production (blue) are shown along with the model-dependent signals of $bW$ fusion (magenta) and $T\bar{t}$ production (orange).}
\label{fig:mumuTT}
\end{figure}

The corresponding Lagrangian is
\begin{equation}
\label{eq:Lag4VLQ}
\begin{aligned}
\mathcal{L} 
& = i \bar{T} \slashed{\partial} T - M_T \bar{T} T - g_s \bar{T} \slashed{G} T
- \frac{2}{3} g'  \bar{T} \slashed{B} T \\
& \quad + \Big( \kappa_W  \frac{g}{\sqrt{2}}\bar{T}_L W^+_\mu\gamma^\mu b_L + \kappa_Z \frac{g}{2c_W}\bar{T}_L Z_\mu\gamma^\mu t_L -\kappa_H \frac{M_T}{v}\bar{T}_L H t_L  + \mathrm{h.c.} \Big)
\end{aligned}
\end{equation}
where $M_T$ is the vector-like mass of the fermion, $\kappa_W$ is the coupling to $Wb$, $\kappa_Z$ is the coupling to $Zt$, and $\kappa_H$ is the coupling $Ht$.  The couplings are considered relative to the Standard Model coupling values.  If the mixing $T-t$ is the only source for $T$ to couple to the SU(2)$_L$ sector, then $\kappa_W=\kappa_Z=\kappa_H$. Throughout the paper, $g_s$ is the QCD coupling constant associated with the gluon field $G$, $g$ is the weak coupling constant, and $g'$ is the hypercharge coupling constant associated with the gauge field $B$. 

After electroweak symmetry breaking, the new heavy fermion $T$ couples to the photon with a charge of $Q=2/3$.  We present the total cross section for $\mu^+\mu^- \to T\bar T$ production in Fig.~\ref{fig:mumuTT} as a function of mass at $\sqrt s=10$ TeV and as a function of center-of-mass energy for $M_T=2$ TeV, separated into $\mu^+\mu^-$ direct annihilation, $\gamma\gamma$ fusion, and QCD production.  The numerical evaluation is performed with \texttt{Madgraph}~\cite{Alwall:2014hca,Frederix:2018nkq} interfaced with the {\tt UFO}~\cite{Degrande:2011ua} model files from Refs.~\cite{delAguila:2000rc,Cacciapaglia:2010vn,Buchkremer:2013bha}.

As shown in Fig.~\ref{fig:mumuTT}, the leading production channel comes from $\mu^+\mu^-$ direct annihilation, with a cross section that scales as $\beta/s$, and reaches the order of femtobarn.  The next-to-leading channel is $\gamma\gamma$ fusion and the rate scales as $(1/M_T^2) \ln^2(\hat s/m_\mu^2)$.  At higher orders, QCD production via quarks and gluons will contribute as shown.  The rate of these production channels is determined entirely from the quantum numbers of the $T$. The vertical axes on the right in Fig.~\ref{fig:mumuTT} show the number of events expected per 10 ab$^{-1}$.

A heavy quark $T$ can also be singly produced through the coupling to $b$ induced by $T-t$ mixing. The production rate depends on the mixing strength which can vary in different models.  The first process of these is $\mu^+ \mu^- \to T \bar{t}$ which proceeds through an $s$-channel $Z$ and consequently is proportional to $\kappa_Z^2/s$.  The second is $bW \to T$ where both the $b$ and $W$ originate from the muon PDF.  The $bW \to T$ cross section scales as $(\kappa_W^2/M_T^2) \ln(\hat s/m_b^2)\ln(\hat s/M_W^2)$.  We show their cross sections in Fig.~\ref{fig:mumuTT}, with the representative values $\kappa_Z=\kappa_W=0.1$.  We see that these channels can be quite competitive with the leading pair production. 

As for the decay of $T$, the leading decay channels are $T\to Wb$, $T \to Zt$, and $T \to Ht$, the widths for which can be derived from Eq.~\eqref{eq:Lag4VLQ}. For $M_T \gg m_W$, the decay partial widths are dominated by the longitudinal gauge bosons and are given by 
\begin{equation}
\Gamma(T\to bW) \approx \kappa_W^2 \frac{ g^2 M_T^3}{64\pi m_W^2},
\quad
\Gamma(T\to tZ)\approx\frac{1}{2}\kappa_Z^2 \frac{ g^2 M_T^3}{64\pi m_W^2},
\quad
\Gamma(T\to tH)\approx\frac{1}{2}\kappa_H^2 \frac{ g^2 M_T^3}{64\pi m_W^2}\ ,
\label{eq:decay} 
\end{equation}
leading to the branching ratios $\mathcal{B}(T\to bW) : \mathcal{B}(T \to tZ) : \mathcal{B}(T \to tH) \approx 1/2:1/4:1/4$ when a univeral coupling $\kappa=\kappa_W=\kappa_Z=\kappa_H$ is used. This is in accordance with the Goldstone Boson Equivalence Theorem~\cite{Lee:1977eg} and the factor of $M_T^2/m_W^2$ in the partial widths reflects the longitudinal gauge boson enhancement.  In fact, if we work in a framework that properly takes into account $T-t$ mixing, the mixing parameters scale as $\kappa \sim m_W/M_T$, which restores the usual perturbative partial width of $\Gamma \sim (g^2/4\pi) M_T$~\cite{Han:2003wu}. 

For a heavy $T$ with a 1 TeV mass, the signal would be spectacular once the center-of-mass energy is above the $T\bar{T}$ production threshold.  In hadronic decays, $3-5$ jets reconstruct the mass $M_T$.  The $T$ and $\bar T$ are produced back-to-back, leaving very little missing energy.  Even including the leptonic decay mode $W^\pm \to \ell^\pm \nu$, the missing energy is only a small fraction of the full event energy.  Furthermore, for $M_T\gg M_W$, the decay products $\ell^\pm \nu$ are highly collimated, providing an additional handle for the kinematic reconstruction. The signal events are essentially background-free in the clean environment of leptonic collisions.  In estimating the discovery potential, we include all decay channels. 

We study the leptonic and hadronic channels of the $T \to W^+b$ decay in some detail, with the parton-level $W^+b \to T$ sample generated with \texttt{Madgraph}.  We impose basic fiducial cuts, motivated by the suggested detector configuration~\cite{MuCoL:2024oxj}
\begin{equation}\label{eq:fid4lj}
p_T^{\ell(j)}>15~\GeV,
\qquad\qquad
|\eta_{\ell(j)}|<2.5,
\qquad\qquad
\Delta R_{\ell j}>0.4.
\end{equation}
In addition, we require the reconstructed invariant mass within
\begin{equation}
0.85\ M_T < m_T^{\rm reco} < 1.15\ M_T,
\end{equation}
to resolve the heavy $T$ resonance and suppress the SM background.  It has been shown that, including the detector energy resolution, this mass window should contain nearly 100\% of the signal events~\cite{Belyaev:2023yym}.  In addition, we assume a $b$-tagging efficiency of 70\% and tagging efficiency of light-flavor jets of 90\%~\cite{Guo:2023jkz}.  The cuts above along with the tagging efficiencies lead to a heavy $T$ reconstruction efficiency of $44-52\%$ which is appropriately weighted between hadronic and leptonic decays.  This efficiency is applied to the cross section to find the estimated event rate.  As an estimation, we adopt the same efficiency for $T\to tZ$ and $T\to tH$.

\begin{figure}
  \centering
  \includegraphics[width=0.55\textwidth]{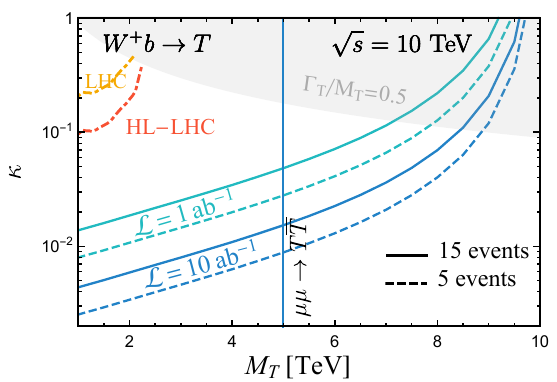}
  \caption{The discovery reach ($N=15$) and exclusion limit ($N=5$) for a color-triplet fermion $T$ as a function of mass $M_T$ and coupling $\kappa$.  The LHC sensitivity and the HL-LHC sensitivity recast are from Ref.~\cite{ATLAS:2024xdc}.}
  \label{fig:tps_reach}
\end{figure}

With the above selection this search becomes effectively background-free and we therefore use the conservative criteria of $N=5$ signal events as the 95\% CL exclusion~\cite{Feldman:1997qc}.  For the discovery contour we consider $N=15$ signal events~\cite{ParticleDataGroup:2024cfk}. As expected, for the model independent $T\bar{T}$ pair production, we find a mass reach that is very close to $\sqrt{s}/2 = 5\; \TeV$. For single $T$ production, the results are shown in Fig.~\ref{fig:tps_reach} with $\mathcal{L}=1~{\rm ab}^{-1}$ and with $\mathcal{L}=10~{\rm ab}^{-1}$.

\subsection{Color-Triplet Scalars: Squarks}

We now consider a color-triplet scalar~\cite{ILC:2013jhg,CLIC:2018fvx}, which are called squarks in SUSY models.  Again, we choose the electroweak and hypercharge quantum numbers to be
\begin{equation*}
\text{color-triplet scalar }(\tilde{q}): 
\quad (3, 1)_{e_q},
\end{equation*}
where $e_q = 2/3$ and $e_q = -1/3$, corresponding to up-type and down-type SUSY partners of the right-handed quarks, respectively.   For our phenomenological presentation, we only consider a single right-handed squark with a charge $e_q = 2/3$, commonly called a ``stop'' $\tilde{t}_R$.  For simplicity we exclude the gluino from contributing to any production or decay channels.

The Lagrangian then reads
\begin{equation}
\begin{aligned}
\mathcal{L} & =
D_\mu \Tilde{t}_R^* D^\mu \Tilde{t}_R -\Tilde{t}_R^* M_{\Tilde{t}_R}^2 \Tilde{t}_R \\
& = \partial_\mu \Tilde{t}_R^* \partial^\mu \Tilde{t}_R - \Tilde{t}_R^* M_{\Tilde{t}_R}^2 \Tilde{t}_R -\frac{2}{3} i g' (\Tilde{t}_R^* \overset{\smash{ \raisebox{-0.2em}{\small{$\leftrightarrow$}}}}{\partial^\mu} \Tilde{t}_R) B_{\mu} +\frac{4}{9}g'^2 B_\mu B^\mu \Tilde{t}_R^* \Tilde{t}_R \\
&\quad \quad - i g_s (\Tilde{t}_R^* T^a \overset{\smash{ \raisebox{-0.2em}{\small{$\leftrightarrow$}}}}{\partial^\mu} \Tilde{t}_R)G^a_\mu + g_s^2\, \Tilde{t}_R^* T^a T^b \Tilde{t}_R\, G^a_\mu\, G^{b\mu} + \frac{4}{3}g'g_s \, \Tilde{t}_R^* T^a \Tilde{t}_R \,G^a_{\mu} B^\mu ,
\label{eq:stop}
\end{aligned}
\end{equation}
where $M_{\tilde{t}_R}$ is the stop mass and $T^a$ are the generators of SU(3) for $a=1,\ldots,8$.
The coupling and structure of the vertices of squark-squark-vector and squark-squark-vector-vector are determined by the squark's quantum numbers and apply for any scalar with those quantum numbers.  As mentioned, we only study $\tilde{t}_R$ which we refer to $\tilde{t}$ hereafter.

\begin{figure} 
  \centering
  \includegraphics[width=0.45\textwidth]{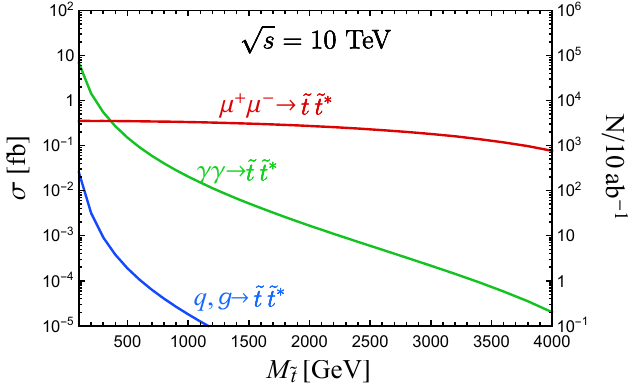}
  \qquad
  \includegraphics[width=0.45\textwidth]{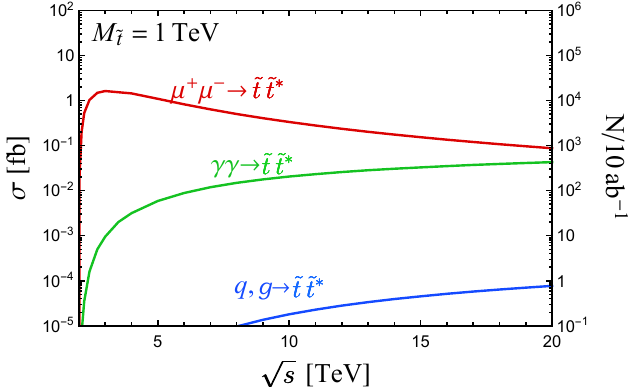}
  \caption{The cross section for the pair production $\tilde{t}\tilde{t}^*$ of a color-triplet scalar $\tilde{t}$ with quantum numbers $(3, 1)_{2/3}$ as a function of mass (left) and as a function of center-of-mass energy (right). The signals of $\mu^+ \mu^-$ annihilation (red), $\gamma\gamma$ fusion (green), and QCD production (blue) are shown.} 
  \label{fig:mumusqsq}
\end{figure}

The electroweak quantum numbers of the $\tilde{t}$ allow for $\mu^+\mu^-$ direct annihilation and $\gamma\gamma$ fusion. The QCD production via quark and gluon scattering is subleading to both of these.  All three production modes are shown in Fig.~\ref{fig:mumusqsq}. The pair production cross section for $\tilde{t}\tilde{t}^*$ is lower than that for $T\bar T$ by about a factor of 4, due to the spin counting and different threshold behavior.  The vertical axes on the right in Fig.~\ref{fig:mumusqsq} show the number of events expected per 10 ab$^{-1}$.

When the electroweakinos, $\chi^\pm$ and $\chi^0$, are lighter than the stops, the stops decay via $\tilde{t} \to t \chi^0 \to bW\chi^0$ or $\tilde{t} \to b\chi^\pm \to bW\chi^0$.  The particle $\chi^0$ is the lightest supersymmetric particle (LSP) and potentially makes up the dark matter of the universe~\cite{Jungman:1995df}.  The two final state neutralinos escape detection and thus lead to missing energy.

This missing energy results in significant consequences for the signal search.  Firstly, we can no longer reconstruct the invariant masses of the heavy particles in the signal selection. Secondly, the missing energy leads to different backgrounds such as $\gamma \gamma \to t \bar t$.  Nevertheless, the signal kinematics are still sufficiently unique to be distinguishable from the SM backgrounds. 

\begin{figure} 
  \centering
  \includegraphics[width=0.55\textwidth]{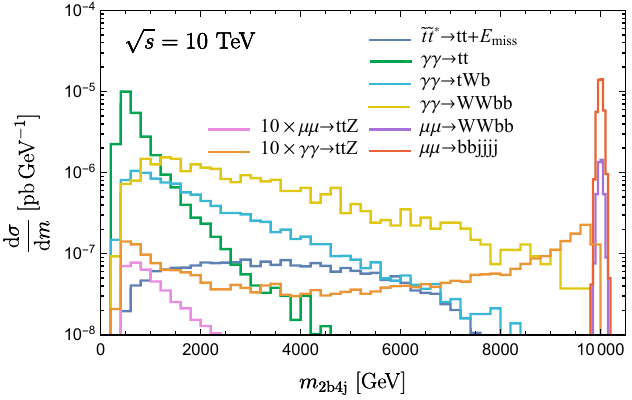}
  \caption{Distribution of $m_{\mathrm{2b4j}}$ for a color-triplet scalar $\tilde{t}$ at $\sqrt{s} = 10~{\rm TeV}$.  The signal point $(M_{\tilde{t}}, M_{\tilde{\chi}^0}) = (2~{\rm TeV}, 830~{\rm GeV})$ is used (dark blue).  All other lines show various backgrounds.  The backgrounds of $\mu\mu \to t\bar{t}Z$ and $\gamma\gamma \to t\bar{t}Z$ are multiplied by $10$ for visibility.}
\label{fig:bkg_m2b4j}
\end{figure}

To estimate the reach for stops, we assume the decay $\tilde{t} \to t \chi^0$ with a branching fraction $\mathcal{B}(\tilde{t}\to t\chi^0\to bW\chi^0)=100\%$.\footnote{Note that the decay through a chargino $\tilde{t} \to \chi^\pm b$ leads to the same final state of $bW\chi^0$.}  We use the fully hadronic decay of the top and anti-top leading to a final state $bbjjjj\slashed{E}_T$.  Motivated by the heavy mass $M_{\tilde{t}}$, we first explore the cluster invariant mass $m_{\mathrm{2b4j}}$ of the visible particles in the final state 
\begin{equation}
m_{\mathrm{2b4j}}^2 = (p_{b1} + p_{b2} + p_{j1} + p_{j2}+ p_{j3}+ p_{j4})^2.  
\end{equation}
The distribution is shown in Fig.~\ref{fig:bkg_m2b4j} for the signal point of $(M_{\tilde{t}}, M_{\tilde{\chi}^0_1}) = (2~{\rm TeV}, 830~{\rm GeV})$ and for the leading backgrounds. For backgrounds from $\mu^+ \mu^-$ annihilation, when there are no on-shell $W$s we label this as $\mu\mu \to bbjjjj$, when there are two on-shell $W$'s, we label this as $\mu \mu \to WWbb$. For $\gamma\gamma$ fusion, similarly, we label the final states as $t$ or $Wb$ according to whether or not there is on-shell $t$.

The signal yields a broad distribution, reaches a plateau around $2 M_{\tilde{t}}$, and is bounded by $\sqrt s -2 M_{\tilde{\chi}^0_1}$. For the $\mu^+ \mu^-$ annihilation backgrounds $\mu\mu \to bbjjjj$ and $\mu\mu \to WWbb$ the $2b$ and $4j$ are the entire final state, so these cluster near $\sqrt{s}$.  For the $\gamma\gamma$ fusion backgrounds $\gamma\gamma \to t\bar{t},tWb,WWbb$, the $\gamma$s are from the collinear radiation and tend to be soft, leading to a distribution at low values of $m_{\mathrm{2b4j}}$. These features provide some discrimination power between the signal and backgrounds. 

The signal usually contains substantial missing transverse momentum, $p_{T\rm miss}$, due to the undetected $\tilde{\chi}^0_1$ particles.  The well-constrained kinematics in leptonic collisions allow for a measurement of the missing mass~\cite{Han:2020uak} $m_{\mathrm{miss}}$ as 
\begin{equation}
m_{\rm miss}^2=p_{\rm miss}^2=\Big(p_{\mu^+} + p_{\mu^-} - \sum_i p_i^{\rm obs}\Big)^2 .
\end{equation}
The four-vectors $p_{\mu^+}$ and $p_{\mu^-}$ are the incoming momenta of the $\mu^+$ and $\mu^-$, respectively, while the summation includes all visible final state particles $i$ with four-vectors $p_i^{\rm obs}$.  For a single invisible particle, this would peak at its on-shell mass as the ``recoil mass'' variable.  For multiple invisible particles, the distribution will have a threshold at the sum of the missing masses, thus distinguishing itself from the SM backgrounds and presenting the possibility to measure the missing particle mass.  

\begin{figure} 
  \centering
  \includegraphics[width=0.45\textwidth]
  {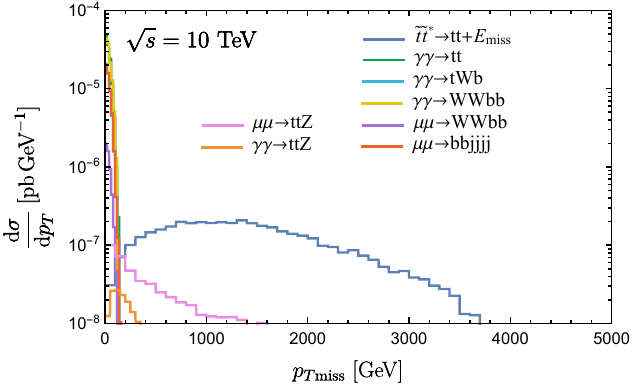}
  \qquad
  \includegraphics[width=0.45\textwidth]
  {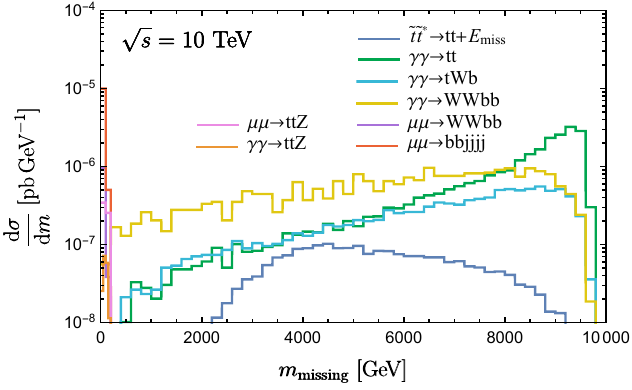}
  \caption{Distribution of 
  missing transverse momentum (left) and missing mass (right) for a color-triplet scalar $\tilde{t}$ at $\sqrt{s} = 10~{\rm TeV}$.  The signal point $(M_{\tilde{t}}, M_{\tilde{\chi}^0}) = (2~{\rm TeV}, 830~{\rm GeV})$ is used (dark blue).  All other lines show various backgrounds.}
  \label{fig:mmissing}
\end{figure}

The distributions of the missing transverse momentum $p_{T,\rm miss}$ and missing mass $m_{\mathrm{miss}}$ are shown in Fig.~\ref{fig:mmissing}, for several signal parameter points.  For $m_{\mathrm{miss}}$, the threshold near $2M_{\tilde{\chi}^0}$ is quite distinctive.  For $p_{T,\rm miss}$, the leading backgrounds of $\mu^+ \mu^- \to bbjjjj$, $\mu^+ \mu^- \to WWbb$, $\gamma\gamma \to t\bar{t}$, $\gamma\gamma \to tWb$, and $\gamma\gamma \to WWbb$ do not have invisible particles so their distributions cluster near zero.

Other backgrounds that include true missing mass could come from a leptonically decaying $W$ or $Z$.  In both cases, the relevant scales of $m_W$ or $m_Z$, respectively, are far below the large missing momentum and missing mass, rendering these backgrounds negligible.  To further purify the signal sample, we may consider the mass $m_t$ reconstruction from the $Wb$ system which will remove the non-top backgrounds.  

\begin{figure} 
  \centering
  \includegraphics[width=0.55\textwidth]{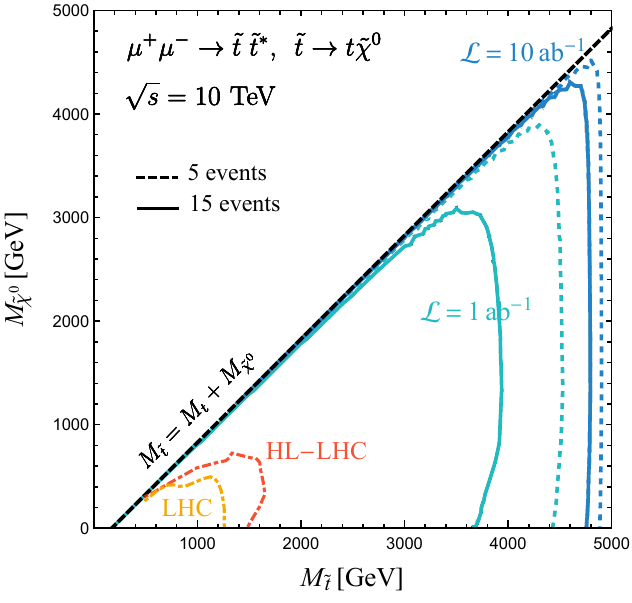}
  \caption{The discovery reach ($N=15$) and exclusion limit ($N=5$) for a color-triplet scalar $\tilde{t}$ as a function of mass $M_{\tilde{t}}$ and neutralino mass $M_{\tilde{\chi}^0}$.  The LHC sensitivity is taken from Ref.~\cite{ATLAS:2024rcx} and the HL-LHC sensitivity is taken from Ref.~\cite{Han:2019grb}.}
  \label{fig:susy_exclusion}
\end{figure}

For signal selection, we adopt the following acceptance cuts
\begin{align} \label{eq:stop_cuts}
p_T^j>15~\GeV, \quad |\eta_j| < 2.5,
\quad
p_{T,\rm miss}>250~\GeV,
\quad
m_{\rm miss}>2~\TeV.
\end{align}
These cuts effectively reduce the main background. Assuming a $b$-tagging efficiency of 70\% and tagging efficiency of light-flavor jets of 90\%~\cite{Guo:2023jkz}, our projections for limits and for discovery are shown in Fig.~\ref{fig:susy_exclusion} as a function of stop mass $M_{\tilde{t}}$ and neutralino mass $M_{\tilde{\chi}^0}$.  An integrated luminosity of $\mathcal{L} = 1~{\rm ab}^{-1}$ leads to a discovery reach around 4 TeV over the bulk of the parameter space.  With $\mathcal{L} = 10~{\rm ab}^{-1}$ the reach starts to saturate at the kinematic limit of 5 TeV. 

\subsection{Color-Sextet Scalars: Diquarks}

A color-sextet diquark transforms as a $\mathbf{{6}}$ under QCD, which may couple to a pair of quarks, depending on its electroweak quantum numbers and spin statistics.  We choose the following representation
\begin{equation*}
\text{color-sextet scalar }(D_6):
\quad
(6, 1)_{4/3},
\end{equation*}
which allows for a coupling of $D_6$ to a pair of $u_R$ quarks. In this way, it is natural to think of the sextet as a scalar diquark.

The color-sextet scalar interaction can be generally written as~\cite{Han:2009ya}
\begin{equation}\label{eq:lag_h6}
\begin{aligned}
\Delta\mathcal{L}= 2\sqrt{2}[\bar K^{ab} D_6 \bar q_a (\lambda_LP_L+\lambda_RP_R)q_b^C + h.c.],  
\end{aligned}    
\end{equation}
where $\bar K^{ab}$ is the coupling matrix where $a$ and $b$ are flavor indices, and $\lambda_{L}$ and $\lambda_{R}$ are the couplings to a pair of left-handed quarks and right-handed quarks, respectively. 

For diquarks, both pair production via SM gauge interactions and single production via $qq$ annihilation are possible.  The pair production rate is determined solely by electroweak gauge couplings of $D_6$ with the leading channel from $\mu^+ \mu^-$ annihilation as shown in Fig.~\ref{fig:DY}, followed by $\gamma\gamma$ fusion shown in Fig.~\ref{fig:VBF} and through QCD shown in Fig.~\ref{fig:qqbQCD}.  The single production depends on the coupling $\lambda_R$, which in a UV model would have a predicted value, but in our parametrization is a free parameter.  The rate can be written as~\cite{Han:2009ya}
\begin{equation}\label{eq:qq2H6}
\sigma(q_1q_2\to D_6)=\int\dd x_1\dd x_2f_{q_1}(x_1)f_{q_2}(x_2)\delta(x_1x_2-M^2_{D_6}/s) \frac{4\pi}{3s}(\lambda_L^2 + \lambda_R^2),
\end{equation}
%

\begin{figure}
  \centering
  \includegraphics[width=0.45\textwidth]{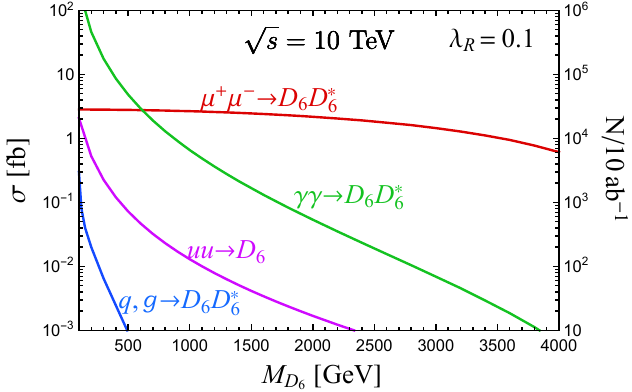}
  \qquad
  \includegraphics[width=0.45\textwidth]{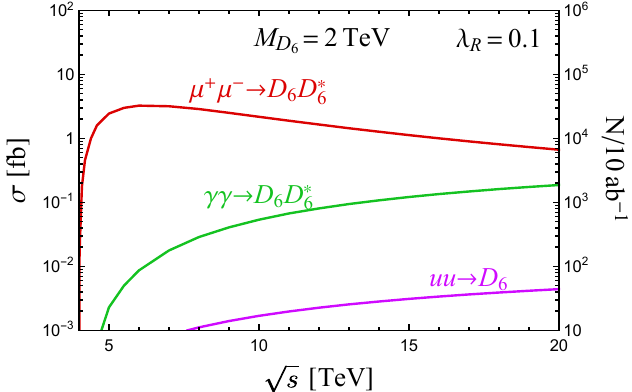}
  \caption{The cross section for the pair production $D_6 D^*_6$ and single production $D_6$ of a color-sextet scalar with quantum numbers $(6, 1)_{4/3}$ as a function of mass (left) and as a function of center-of-mass energy (right).}
  \label{fig:sextet}
\end{figure}

To satisfy the stringent constraints from flavor physics, we choose the flavor structure $\bar{K}^{uu}=1$ and zero for other elements of $\bar{K}^{ab}$ in Eq.~\eqref{eq:lag_h6}.  We use the Yukawa couplings to right-handed quarks of $\lambda_R=0.1$ and since there is no coupling to left-handed quarks, effectively we have $\lambda_L=0$.  The cross sections for pair production and single production are shown in Fig.~\ref{fig:sextet}.

Similar to the production of other heavy states via electroweak interactions, the cross sections for $\mu^+\mu^- \to D_6 D_6^*$ and $\gamma\gamma \to D_6 D_6^*$ can be sizable, while the contributions from QCD interactions, via the $u$ and $g$ content of the muon, are smaller by about two orders of magnitude.  At hadron colliders, like the LHC, QCD production is dominant and leads to very strong bounds on diquark masses and couplings~\cite{CMS:2019gwf}, as shown in Table~\ref{tab:Model2}.  Since the LHC bounds already surpass 7 TeV, we will not make detailed sensitivity estimates for $D_6$ at a muon collider.

\section{Colored Leptons: Leptoquarks and Leptogluons}
\label{sec:fusion}

In the Standard Model, leptons only interact through the electroweak and Yukawa interactions.  Many fundamental puzzles, such as the existence of three generations of fermions, the quark-lepton flavor structure, as well as possible grand unification suggest the possibility of new mechanisms beyond the Standard Model.  Hypothetical colored leptons, such as leptoquarks, are such particles predicted by many BSM theories of unification, like the Pati-Salam model~\cite{Pati:1974yy} or superstring models~\cite{Dobado:1987pj}.  Similarly, the leptogluon, another colored lepton, is predicted in composite models and corresponds to a bound state of at least some colored constituents~\cite{Fritzsch:1981zh,Harari:1982xy}.

Colored leptons are intriguing at muon colliders because single production occurs via the muon from one beam and via a colored parton of the other beam.  The corresponding production mechanism includes QCD scattering represented by $\sigma_{\rm QCD}$ in Eq.~\eqref{eq:power}, which scales as $\alpha^n \alpha_s^m$, where the indices $n,m$ denote the QED/EW and QCD interaction orders, respectively, including both collinear splittings and hard scattering in Fig.~\ref{fig:qqbQCD} distinct from the cases in Sec.~\ref{sec:EWprod}.  In this section, we present the production cross section of colored leptons and correspondingly explore search strategies. 

\subsection{Scalar Leptoquarks}
\label{sec:leptoquarks}

\begin{table}
    \centering
    \begin{tabular}{c|c|c|c|c|c|c}
    \hline
    ~~$S_{\ell q}$~~ & ~$SU(2)_L$~  & ~$U(1)_Y$~ & $T_3$ & Charge $Q$ & Interactions & Existing bounds  \\
    \hline
    $S_1^{1/3}$ & 1 & 1/3 & 0 & 1/3 &  $\bar{q}_L^c\ell_L$, $\bar{u}_R^ce_R$ & $1.4-1.7$ TeV \\
    $S_1^{4/3}$ & 1 & 4/3 & 0 & 4/3 &  $\bar{d}_R^c e_R$ & 1.8 TeV  \\    
    $S_{2,7/6}^{5/3}$, $S_{2,7/6}^{2/3}$ & 2& 7/6 & $1/2, -1/2$ & 5/3, 2/3 & $\bar{q}_Le_R$, $\bar{u}_R\ell_L$ & $1.9-2.0$ TeV\\
    $S_{2,1/6}^{2/3}$, $S_{2,1/6}^{-1/3}$ & 2& 1/6 & $1/2, -1/2$ & $2/3,-1/3$ & $\bar{d}_L\ell_L$ & 1.7 TeV\\
    $S_{3}^{4/3}$, $S_{3}^{1/3}$, $S_{3}^{-2/3}$ & 3 & 1/3 & $1,0,-1$ &  $4/3,1/3,-2/3$ &  $\bar{q}_L^c\ell_L$ & $1.8\; \TeV$ \\
    \hline
    \end{tabular}
    \caption{Summary of the scalar leptoquark $S_{\ell q}$ with their quantum numbers, their allowed interactions, and the existing bounds taken from Refs.~\cite{Bhaskar:2023ftn,CMS:2024bnj}.  When there is more than one allowed interaction the bound is a range that depends on the branching fraction of each interaction.}
    \label{tab:SLQ}
\end{table}

As their name suggests, leptoquarks couple directly to leptons and quarks, and can be either scalars or vectors.  For a massive vector leptoquark, the production mode can receive an anomalous enhancement due to a missing cancellation of the longitudinal mode, depending on the specific UV completion for its mass generation.  A specific example of a color-triplet vector leptoquark with electroweak quantum numbers $1_{2/3}$ that couples to the second and third generations is explored in Ref.~\cite{Asadi:2021gah}.  Detailed studies on other representations and flavor couplings are left for future work.  In this work, we focus on scalar leptoquarks (SLQ).

Following the Particle Data Group~\cite{ParticleDataGroup:2024cfk}, we list all possible scalar leptoquarks, along with their corresponding quantum numbers and allowed interactions in Table~\ref{tab:SLQ}.  We use the notation of $S_{\ell q}$ to refer to leptoquarks in general and notate specific leptoquarks, for example, as $S_{2,1/6}^{2/3}$, where the subscript lists the SU(2)$_L$ representation and the hypercharge and the superscript lists the electric charge.  For SU(2)$_L$ singlets and triplets the hypercharge subscript is omitted.  The Lagrangian of an SLQ is
\begin{equation}\label{eq:Lkin}
\mathcal{L}=D_\mu S_{\ell q}^*D^\mu S_{\ell q}-M_{S}S_{\ell q}^*S_{\ell q}-\mathcal{L}_Y,
\end{equation}
where the Yukawa-like interactions are
\begin{equation}
\begin{aligned}\label{eq:LY4SLQ}
\mathcal{L}_Y
& =S_{1}^{1/3}\left(\lambda_{R}^{1/3}\bar{u}_R^c e_R+\lambda_L^{1/3}\bar{q}_L^c\ell_L\right)+
\lambda^{4/3}S_{1}^{4/3}\bar{d}_R^ce_R+
\lambda_{3}\bar{q}_L^c\frac{\sigma^{a}}{2}\ell_LS_3^{a}\\
& \qquad
+S_{2,7/6}\left(\lambda_2^{qe}\bar{q}_Le_R+\lambda_2^{u\ell}\bar{u}_Ri\sigma_2\ell_L\right)
+\lambda_{2,1/6}S_2^{1/6}\bar{d}_R\ell_L + \textrm{h.c.}
\end{aligned}
\end{equation}
For simplicity, we only consider couplings to the $u$ quark and to the muon.

\begin{figure}
  \centering 
  \subfigure[]{\includegraphics[width=0.23\linewidth]{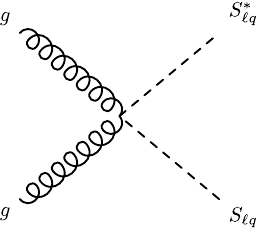}\label{fig:gg2LQLQ}}     
  \subfigure[]{\includegraphics[width=0.23\linewidth]{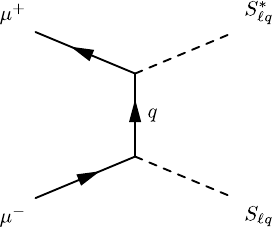}\label{fig:mumu2LQLQ}} 
  \subfigure[]{\includegraphics[width=0.27\linewidth]{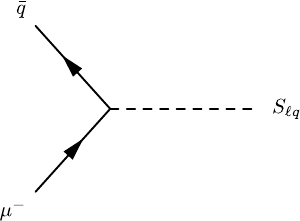}\label{fig:qmu2LQ}}
  \subfigure[]{\includegraphics[width=0.23\linewidth]{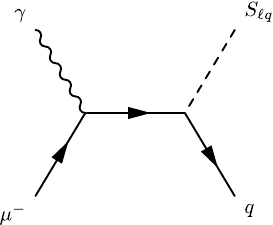}\label{fig:amu2qLQ}}   
  \caption{Feynman diagrams for the pair production of scalar leptoquarks from (a) gluon-gluon fusion and (b) $t$-channel quark mediation, and for the single production of a scalar leptoquark from (c) quark-muon annihilation and (d) in Compton-like scattering.}
  \label{fig:feynLQ}
\end{figure}

As SLQs carry both electric charge and color, they are pair produced through direct annihilation $\mu^+ \mu^-\to \gamma^*,Z^* \to S_{\ell q}S_{\ell q}^*$ in Fig.~\ref{fig:DY}, photon-photon fusion $\gamma\gamma\to S_{\ell q}S_{\ell q}^*$ in Fig.~\ref{fig:VBF}, in quark-antiquark annihilation $q\bar{q}\to S_{\ell q}S_{\ell q}^*$ in Fig.~\ref{fig:qqbQCD}, and in gluon gluon fusion in Fig.~\ref{fig:gg2LQLQ}.  From the interactions in Eq.~\eqref{eq:LY4SLQ}, the SLQs are also pair produced through the mediation of a $t$-channel quark in Fig.~\ref{fig:mumu2LQLQ}.  An SLQ is single produced by quark-muon fusion $q\mu\to S_{\ell q}$ in Fig.~\ref{fig:qmu2LQ} and by Compton-like scattering $\gamma\mu\to q S_{\ell q}$ in Fig.~\ref{fig:amu2qLQ}.

\begin{figure} 
  \centering
  \includegraphics[width=0.45\textwidth]{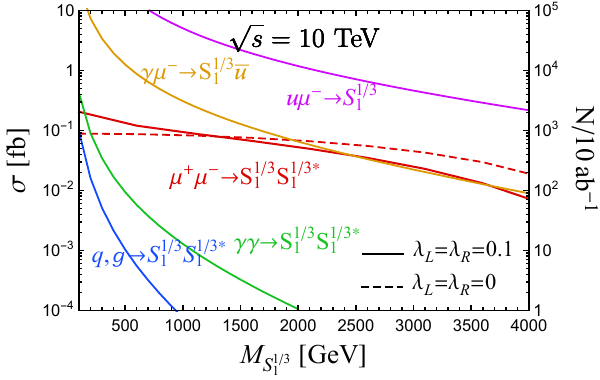}
  \qquad
  \includegraphics[width=0.45\textwidth]{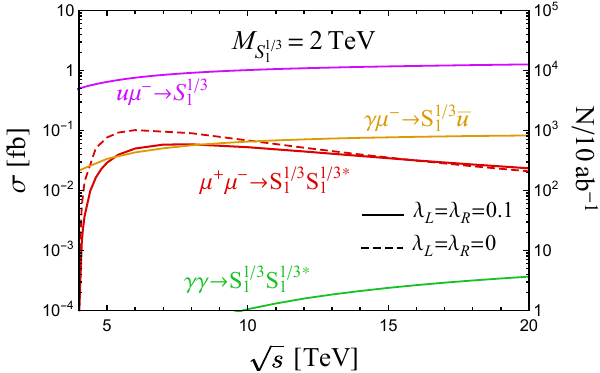}
  \caption{The cross section for the pair production $S_1^{1/3} S_1^{1/3*}$ and single production $S_1^{1/3}$ of a scalar leptoquark $S_1^{1/3}$ as a function of mass (left) and as a function of center-of-mass energy (right). }
  \label{fig:xsLQ}
\end{figure}

\begin{figure}
  \centering
  \includegraphics[width=0.55\linewidth]{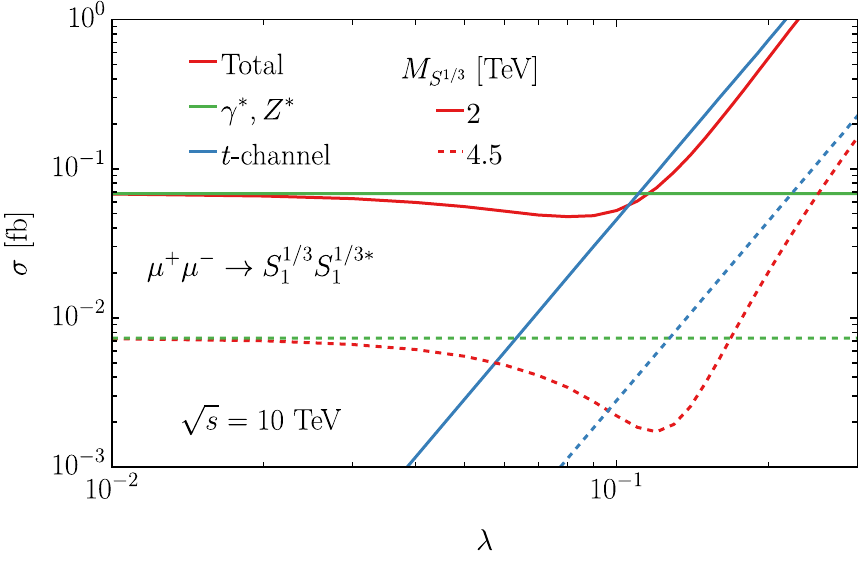}
  \caption{The cross section for the pair production $S_1^{1/3} S_1^{1/3*}$ of a leptoquark $S_1^{1/3}$ as a function of Yukawa coupling.  The masses of $S_1^{1/3}$ shown are 2 TeV (solid) and 4.5 TeV (dashed).  The total (red), the $\gamma^*,Z^*$ mediated (green), and the $t$-channel quark-mediated (blue) cross sections are shown.}
  \label{fig:xsLQlambda}
\end{figure}

We show each of these production channels individually for the $S_1^{1/3}$ leptoquark in Fig.~\ref{fig:xsLQ}.  This SLQ has two possible couplings $\lambda_L$ and $\lambda_R$, both of which lead to a vertex of $u\mu S_1^{1/3}$.  To reduce our parameter space we will set these couplings equal to a common value $\lambda = \lambda_L = \lambda_R$.  The cross sections with only gauge couplings and $\lambda_L = 0$ and $\lambda_R = 0$ are shown with a dashed line.  The cross sections with $\lambda_L = 0.1$ and $\lambda_R = 0.1$ are shown with solid lines.  When the Yukawa-like couplings are non-zero, several different production channels open up and the direct annihilation of $\mu^+ \mu^- \to S_1^{1/3} S_1^{1/3*}$ changes due to interference with the $t$-channel quark diagram as seen in Fig.~\ref{fig:mumu2LQLQ}.

The behavior of the total cross section as the Yukawa-like coupling changes is shown in Fig.~\ref{fig:xsLQlambda}.  When the Yukawa-like coupling is small, $\lambda <0.05$, the cross section is nearly constant as it is dominated by $\mu^+ \mu^-\to \gamma^*,Z^* \to S_{\ell q}S_{\ell q}^*$.  On the other hand, the cross section for the $t$-channel quark diagram scales as the fourth power of the Yukawa coupling, and thus dominates the pair production for $\lambda >0.2$. It is particularly interesting to note that there is significant destructive interference between these sets of diagrams. If a signal is observed in this region, the interference effect would serve as the best measurement for the Yukawa coupling. 

In comparison, the single production of an SLQ necessarily involves the Yukawa-like interactions.  The quark-muon fusion cross section is
\begin{equation}
\sigma(q\mu\to S_{\ell q})=
\int\dd x f_{q}(x,Q^2)\frac{\pi}{2}(\lambda_L^2 + \lambda_R^2)\,\delta(\hat{s}-M_{S}^2)
=\frac{\pi}{2s}(\lambda_L^2 + \lambda_R^2)\, f_{q}\!\left(M_S^2/s,M_S^2\right),
\end{equation}
where $\hat{s}=xs$ and we choose the factorization scale as $Q^2=M_S^2$.  The results are shown in Fig.~\ref{fig:xsLQ}. Due to the single particle phase space, the production rate is potentially higher than the other channels.
The next-to-most competitive channel is single SLQ production associated with a quark via the Compton-like process $\gamma \mu \to S_{\ell q}\, u$ in the $s$-channel, as shown in Fig.~\ref{fig:amu2qLQ}. Although it benefits from the single massive SLQ in the final state, the $2\to 1$ fusion process $u \mu \to S_{\ell q}$ is more dominant by about an order of magnitude, as seen in Fig.~\ref{fig:xsLQ}, owing to the phase space and the collinear quark PDF. 

At a high-energy muon collider, once the heavy SLQ is produced it uniquely decays into a lepton and a jet. The lepton-jet final state reconstructs its resonance mass, leading to a spectacular signal.  We apply the fiducial cuts from Eq.~\eqref{eq:fid4lj} to our simulated sample.  The cut efficiency for SLQ pair production is close to 100\%.  The cut efficiency for SLQ single production varies from 19\% when $M_S =$ 1 TeV, up to 95\% when $M_S =$ 5 TeV.  The rather low signal efficiency at low masses is due to the large boost which reduces the opening angle between the jet and lepton and the high rapidity $\ln(M_S^2/s) \approx 5$ when $M_S =$ 1 TeV, compared to $\ln(M_S^2/s) \approx 1.4$ when $M_S =$ 5 TeV.

\begin{figure}
  \centering
  \subfigure[]{\includegraphics[width=0.45\linewidth]{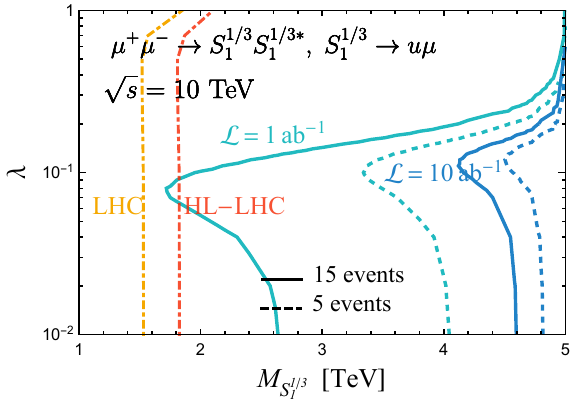}\label{fig:S1S1}}
  \qquad
  \subfigure[]{\includegraphics[width=0.45\linewidth]{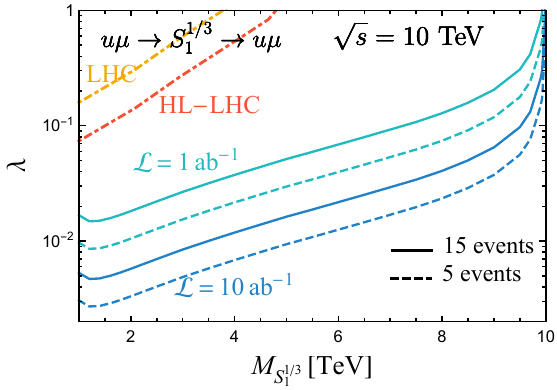}\label{fig:S1}}
  \caption{The discovery reach ($N=15$) and exclusion limit ($N=5$) for a leptoquark $S_1^{1/3}$ in pair production (left) and in single production (right) as a function of mass and coupling.  For pair production the LHC and HL-LHC limits are recast from Ref.~\cite{CMS:2024bnj}.  For single production the LHC and HL-LHC limits are recast from Refs.~\cite{Campana,Bhaskar:2023ftn}.}
  \label{fig:S1_exclusion}
\end{figure}

In the clean lepton collider environment, the invariant mass of the lepton-jet pair can be well constructed, with a peak around the SLQ mass, $M_{\ell j} \approx M_S$.  In our results, we apply a 90\% reconstruction efficiency to jets and a 90\% reconstruction efficiency to leptons.   At such a high lepton-jet invariant mass, we expect that the SM background can be safely neglected.   Our exclusion and discovery projections use the criteria of $N=5$ and $N=15$ events, respectively, with the results shown in Fig.~\ref{fig:S1_exclusion}.

For our sensitivity studies, we focus on the two leading channels $\mu^+\mu^- \to S_{\ell q}S_{\ell q}^*$ and $u\mu \to S_{\ell q}$.
For SLQ pair production in Fig.~\ref{fig:S1S1}, there are two asymptotic limits.  When the Yukawa-like coupling is negligible, the SLQ pair production cross section depends only on the SLQ mass.  As a consequence, we obtain a model-independent sensitivity on the SLQ mass, with the discovery and exclusion limits of 2.6 (4) TeV and 4.6 (4.8) TeV based on an integrated luminosity of $\mathcal{L}=1~(10)~\textrm{ab}^{-1}$.

On the other hand, when the Yukawa-like coupling is large, the $t$-channel diagram in Fig.~\ref{fig:mumu2LQLQ} dominates the production, and leads to a dependence on the fourth power of coupling for the cross section as shown in Fig.~\ref{fig:xsLQlambda}.  In this regime, the sensitivity depends strongly on the coupling.  In between these two regions, we obtain an overturn for the coupling sensitivity, due to the destructive interference from the diagram in Fig.~\ref{fig:mumu2LQLQ}.

In comparison, the single production sensitivity in Fig.~\ref{fig:S1} is only mediated through the Yukawa-like interaction from the diagrams in Fig.~\ref{fig:qmu2LQ} and Fig.~\ref{fig:amu2qLQ}.  The single production limits extend to small couplings and to larger SLQ masses because the single production phase space is larger.

In Fig.~\ref{fig:S1_exclusion}, we also include the existing bounds from the LHC measurements and the projection of the HL-LHC for comparison.  The current LHC exclusion for SLQ pair production is recast from the CMS 13 TeV measurement~\cite{CMS:2024bnj}, while the HL-LHC projection is obtained by rescaling the corresponding luminosity to $3~\textrm{fb}^{-1}$.  The model-independent bound $M_{S}\gtrsim 1.8~\TeV$ presented by CMS~\cite{CMS:2024bnj} assumes a branching ratio of 1 and only applies to the $\lambda_L=0$ limit.  Our recast restores the coupling dependence and uses a branching ratio of $2/3$ corresponding to our benchmark of $\lambda_L = \lambda_R$.
When the coupling becomes large, $\gtrsim 1$, the bound on $M_S$ becomes stronger due to a $t$-channel diagram that is similar to the one in Fig.~\ref{fig:mumu2LQLQ}.  In comparison, the LHC and HL-LHC single production limits are similarly recast from Refs.~\cite{Campana,Bhaskar:2023ftn}.  In both the pair production and single production scenarios, a 10 TeV muon collider has a much stronger sensitivity than the HL-LHC.

\begin{figure}
  \centering
  \includegraphics[width=0.45\linewidth]{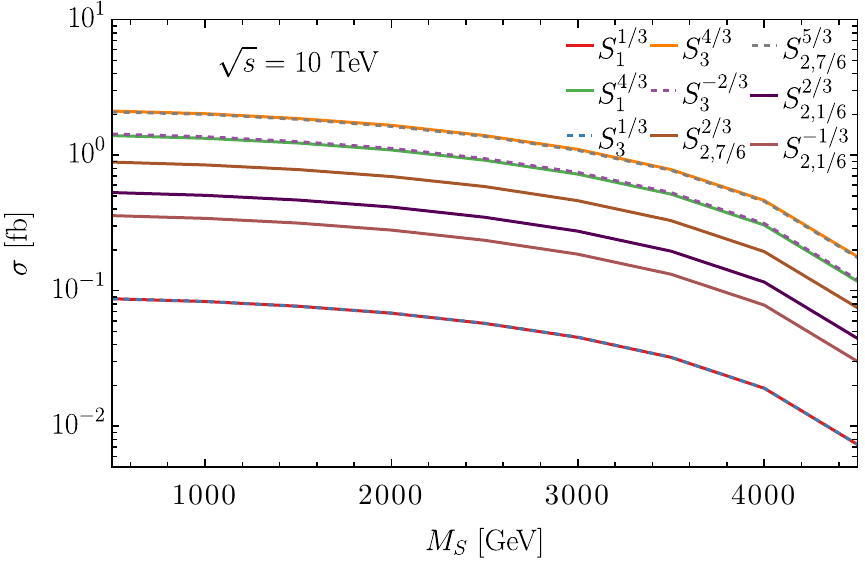}
  \qquad
  \includegraphics[width=0.45\linewidth]{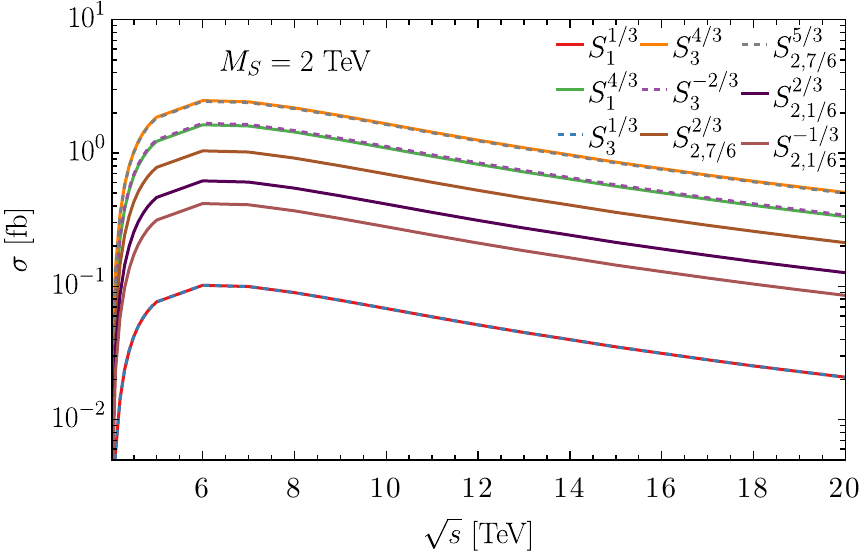}    
  \caption{The cross section for the pair production $S_{\ell_q} S_{\ell_q}^*$ of a scalar leptoquark $S_{\ell_q}$ as a function of mass (left) and as a function of center-of-mass energy (right).  The Yukawa-like coupling is set to $\lambda=0$.}
  \label{fig:SLQ}
\end{figure}

Note that the reach for other SLQ particles, listed in Table~\ref{tab:SLQ}, can be explored in a similar way.  In the absence of the Yukawa-like interaction, the SLQ can only be produced in pairs through the electroweak scattering.  When $\sqrt{s}\gg M_Z$, the full phase-space cross section for SLQ pair production is given by~\cite{Ruckl:1997ex}
\begin{equation}\label{eq:xsecSLQ}
    \sigma = \frac{\pi \alpha^2 \beta^3}{2s} \sum_{\substack{a=L,R\\[0.5mm] V=\gamma,Z}} \big(Q^V_a(\mu) Q^V(S_{\ell q})\big)^2,
\end{equation}
where $\beta=\sqrt{1-4M_S^2/s}$ and
\begin{equation}
 \begin{aligned}
    & Q_{L,R}^\gamma(\mu) =1, 
    \qquad\qquad
    Q_L^Z(\mu) = \frac{-1/2+s_W^2}{s_W c_W}, 
    \qquad\qquad
    Q^Z_R(\mu)=\frac{s_W}{c_W}, \\
    & Q^\gamma (S_{\ell q}) = Q,
    \qquad \qquad
    Q^Z(S_{\ell q}) = \frac{T_3-Q s_W^2}{s_W c_W},
\end{aligned}   
\end{equation}
where $s_W$ and $c_W$ are the sine and cosine of the weak-mixing angle, respectively. 

In Fig.~\ref{fig:SLQ}, we show the SLQ pair production cross section as a function of the SLQ mass at a 10 TeV muon collider and as a function of the collider energy with the SLQ mass $M_{S}=2~\TeV$ for a variety of SLQ quantum numbers.  The general behavior as a function of mass is due to phase space which leads to similar behavior for all SLQs, while the collider energy dependence comes from the $s$-channel $1/s$ behavior in Eq.~(\ref{eq:xsecSLQ}).  The overall size of the cross section only depends on the electroweak charges, shown in Table~\ref{tab:SLQ}.  The model-independent exclusion limits can be recast accordingly.  As shown in Fig.~\ref{fig:xsLQlambda}, a sizable Yukawa-like coupling would change the cross section prediction. This provides the opportunity to measure the Yukawa-like coupling.  Once the signal is observed and the $S_{\ell q}$ state is reconstructed from the decay products, one can compare the predicted cross section to extract the size of the Yukawa-like coupling.

\subsection{Fermionic Leptogluons}

As another type of colored leptons, leptogluons (LG), are predicted by some composite models and carry non-vanishing lepton number and color charges~\cite{Harari:1982xy}. 
Interacting with leptons and gluons directly, LGs are color-octet fermions, with the representation
\begin{equation*}
\text{color-octet fermion }(\ell_g): 
\qquad 
(8, 2)_{-1/2}.
\end{equation*}
The effective Lagrangian is~\cite{Goncalves-Netto:2013nla,Almeida:2022udp}
\begin{equation}\label{eq:L4LG}
\begin{aligned}
    \mathcal{L}=&\bar \ell_g (i \slashed D -M_{\ell_g})\ell_g +\frac{g_s}{2 \Lambda}\bar \ell_g^a \sigma^{\mu\nu}G_{\mu\nu}^a(a_LP_L+a_RP_R)\ell, 
\end{aligned}    
\end{equation}
where $G^a_{\mu\nu}=\partial_\mu G^a_\nu-\partial_\nu G^a_\mu + g_s f^{abc}G^b_{\mu}G^c_{\nu}$ and $\ell$ represents any charged lepton, although only the muon is relevant in this work.

Requiring the interaction of a leptogluon with a lepton and gluon fixes the possible hypercharges to $-1/2$ and $1$ for a SU(2)$_L$ doublet and singlet, respectively.  We choose the SU(2)$_L$ doublet which means only $a_L$ is relevant in our work.  The LG, which couples to the charged lepton, has an electric charge of $\pm 1$.  By gauge invariance, there is a neutral state, that couples to neutrinos and gluons, but we leave the study of such a state for future work.

\begin{figure}
  \centering
  \subfigure[]{\includegraphics[width=0.25\linewidth]{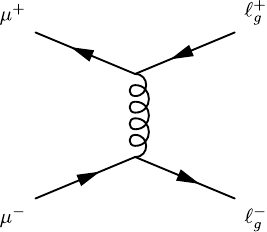}\label{fig:mumu2LGLG}}
  \qquad\qquad
  \subfigure[]{\includegraphics[width=0.25\linewidth]{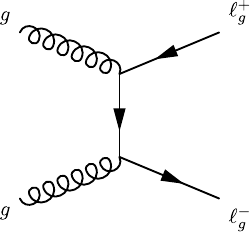}\label{fig:gg2LGLG}}
  \caption{Feynman diagrams for the pair production of fermionic leptogluons through (a) $t$-channel $\mu^+\mu^-$ scattering and (b) gluon-gluon fusion.}
  \label{fig:feynLGLG}
\end{figure}

\begin{figure}
  \centering
  \subfigure[]{\includegraphics[width=0.24\linewidth]{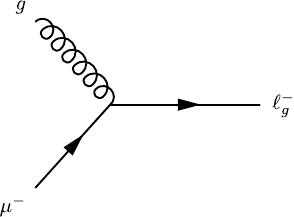}\label{fig:mug2LG}}
  \subfigure[]{\includegraphics[width=0.24\linewidth]{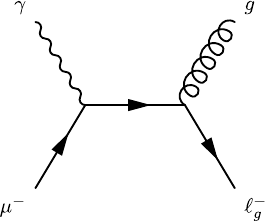}\label{fig:amu2gLG}}
  \subfigure[]{\includegraphics[width=0.24\linewidth]{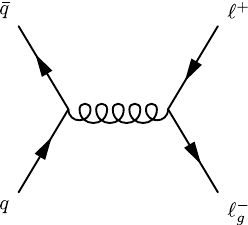}\label{fig:qqb2muLG}}
  \subfigure[]{\includegraphics[width=0.24\linewidth]{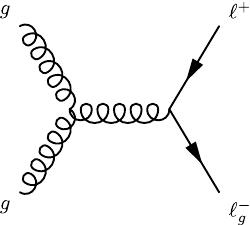}\label{fig:gg2muLG}}
  \caption{Feynman diagrams for the single production of a fermionic leptogluon through (a) gluon-muon fusion, (b) Compton-like scattering, (c) quark-antiquark annihilation, and (d) gluon-gluon fusion.}
  \label{fig:feynLG}
\end{figure}

\begin{figure} 
  \centering
  \includegraphics[width=0.45\textwidth]{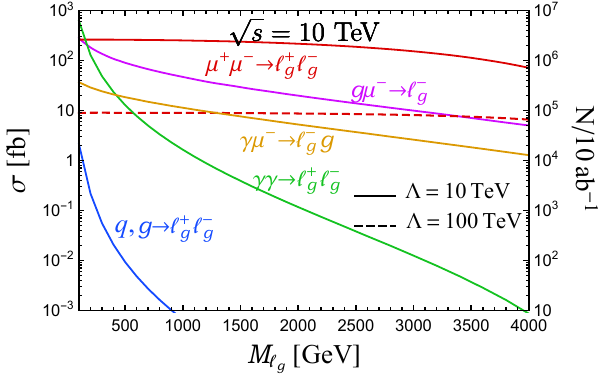}
  \qquad
  \includegraphics[width=0.45\textwidth]{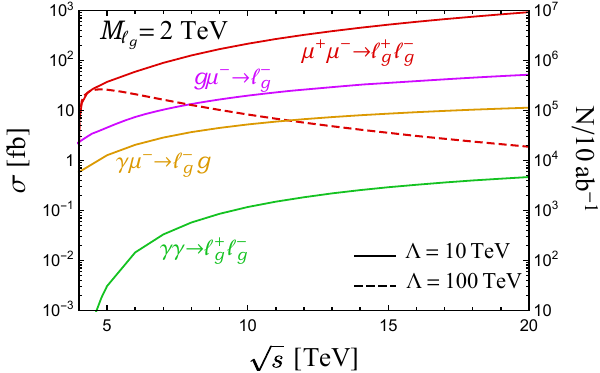}
  \caption{The cross section for the pair production $\ell_g^+ \ell_g^-$ and single production $\ell_g^\pm$ of a leptogluon $\ell_g^\pm$ as a function of mass (left) and as a function of center-of-mass energy (right) with a coupling $a_L=1$ and for the cutoff scales $\Lambda = 10~{\rm TeV}$ (solid).  For $\Lambda = 100~{\rm TeV}$ only $\mu^+ \mu^- \to \ell_g^+ \ell_g^-$ is shown (dashed).}
  \label{fig:xsLG}
\end{figure}

As the leptogluon carries both color and electric charges, for pair production, there are contributions from muon annihilation $\mu^+\mu^-\to\gamma^*/Z^*\to\ell_g^+\ell_g^-$ in Fig.~\ref{fig:DY}, from photon-photon fusion $\gamma\gamma\to\ell_g^+\ell_g^-$ in Fig.~\ref{fig:VBF}, and from quark-antiquark annihilation in Fig.~\ref{fig:qqbQCD}.  In addition, there are contributions from the $t$-channel $\mu^+\mu^-$ scattering in Fig.~\ref{fig:mumu2LGLG} and from gluon-gluon fusion in Fig.~\ref{fig:gg2LGLG}. For single LG production, we have muon-gluon fusion in Fig.~\ref{fig:mug2LG}, Compton-like scattering in Fig.~\ref{fig:amu2gLG}, quark-antiquark annihilation $q\bar{q}\to\ell_g^+\mu^-$ in Fig.~\ref{fig:qqb2muLG}, and gluon-gluon fusion in Fig.~\ref{fig:gg2muLG}.
Similar to Refs.~\cite{Mandal:2016csb,Almeida:2022udp}, we take $a_L=1$ ($a_R$ is absent in our model). In Fig.~\ref{fig:xsLG}, we present the cross sections of pair production and single production with the cutoff scale $\Lambda=10~{\rm TeV}$ for all processes (solid) and for $\Lambda=100~{\rm TeV}$ we show only $\mu^+ \mu^- \to \ell_g^+ \ell_g^-$ (dashed) since all other processes are either identical or simple rescalings.

For single leptogluon production, the gluon-muon fusion cross section behaves as  
\begin{equation}
\sigma(g\mu\to\ell_g)= \int\dd x f_{g}(x,Q^2)\, 2\pi g_s^2 \frac{a_L^2 M_{\ell_g}^2}{\Lambda^2}\,\delta(\hat{s}-M_{\ell_g}^2)= \frac{2\pi g_s^2}{s}  \frac{a_L^2 M_{\ell_g}^2}{\Lambda^2} f_{g}\!\left(M_{\ell_g}^2/s,M_{\ell_g}^2\right),
\end{equation}
where the scale can be chosen as $Q^2=M_{\ell_g}^2$.  For leptogluon pair production, we also obtain a strong dependence on the lepton-gluon-leptogluon coupling through the $t$-channel diagram in Fig.~\ref{fig:mumu2LGLG}, with a clear transition from where $\mu^+\mu^-$ annihilation dominates, for small $a_L/\Lambda$, to where the $t$-channel process dominates, for large $a_L/\Lambda$, with destructive interference in between, similar to Fig.~\ref{fig:xsLQlambda}.

\begin{figure}
  \centering
  \includegraphics[width=0.55\textwidth]{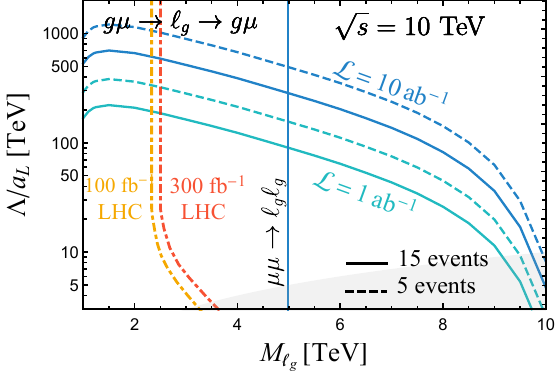}
  \caption{The discovery reach ($N=15$) and exclusion limit ($N=5$) for a fermionic leptogluon $\ell_g$ as a function of mass $M_{\ell_g}$ and scale $\Lambda/a_L$.  
  The shaded region indicates $M_{\ell_g}>\Lambda/a_L$ where effective Lagrangian in Eq.~\eqref{eq:L4LG} is no longer valid.   The LHC sensitivity is from Ref.~\cite{Mandal:2016csb}.}
  \label{fig:lg_exclusion}
\end{figure}

In our simulation, we use fiducial cuts from Eq.~\eqref{eq:fid4lj}, 
which gives the acceptance ranging from $20\%$ at $M_{\ell_g} = 1~{\rm TeV}$ to $95\%$ at $M_{\ell_g} = 5~{\rm TeV}$.  The low efficiency at lower masses is due to the higher rapidity of the final state particles.  In pair production, the efficiency is about 100\%.  We apply a 90\%  reconstruction efficiency to leptons and a 90\% reconstruction efficiency to jets.  With the background-free assumption and the criterion of 5 and 15 events as the exclusion and discovery limits, we present the projected sensitivity of a 10 TeV muon collider in Fig.~\ref{fig:lg_exclusion}. 

From leptogluon pair production, we obtain an exclusion limit for the leptogluon mass of about $M_{\ell_g}>4.99~\TeV$, with little dependence on the lepton-gluon-leptogluon coupling.  This limit saturates the kinematically available phase space because even without any lepton-gluon-leptogluon coupling the pair production rate is already very large.

In comparison, leptogluon single production effectively probes the lepton-gluon-leptogluon coupling $a_L/\Lambda$, as shown in Fig.~\ref{fig:lg_exclusion}.  The existing LHC bound~\cite{Mandal:2016csb,Almeida:2022udp}, which includes both single and pair production is shown in yellow and sets a limit just below 2.5 TeV.  The HL-LHC is only expected to marginally extend the mass range.  A 10 TeV muon collider extends the reach model-independently to 5 TeV, and model-dependently as high as $8-9~{\rm TeV}$.

\section{QCD Scattering for Color Multiplets}
\label{sec:QCD}

In the preceding sections, we have explored colored particles that carry electroweak charge at high-energy muon colliders.  The dominant production mechanism for electroweakly-charged particles comes from the electroweak scattering, including pair production through the Drell-Yan-like $\mu^+\mu^-$ annihilation (including interference) and single production through fusion, such as $Wb\to T$, $\ell q \to S_{\ell q}$, and $\ell g \to \ell_g$.  In this section, we study colored particles that are neutral under the electroweak force and can therefore only be produced through the QCD sector. 

The standard-model quarks are color triplets, \textbf{3}, under $SU(3)_C$. 
A quark and antiquark system can be decomposed as $\textbf{3}\otimes\bar{\textbf{3}}=\textbf{8}\oplus\textbf{1}$, which leads to a color octet channel and a singlet channel.  Color octet resonances arise in many BSM scenarios, such as technicolor models~\cite{Hill:2002ap,Chivukula:1995dt}, universal extra dimensions~\cite{Dobrescu:2007xf,Dobrescu:2007yp}, axigluons~\cite{Frampton:1987dn,Bagger:1987fz}, colorons~\cite{Hill:1993hs,Chivukula:1996yr}, and Kaluza-Klein (KK) gluons~\cite{Dicus:2000hm}.
In this work, we consider color octets, including scalars, fermions, and vectors.

\subsection{Color-Octet Scalars: Techni-pions}

First, we consider a color-octet scalar with the gauge quantum numbers
\begin{equation*}
\text{color-octet scalar }(S_8):
\qquad
(8, 1)_{0}.
\end{equation*}
Such a particle is present in technicolor models as ``techni-pions'' \cite{Hill:2002ap} and in universal extra-dimensional models~\cite{Dobrescu:2007xf,Dobrescu:2007yp}.  The corresponding Lagrangian is~\cite{Idilbi:2009cc,Cao:2021qqt,Han:2010rf}
\begin{equation}\label{eq:S8}
\begin{aligned}
\mathcal{L}=
\frac{1}{2}D_\mu S_8^a D^\mu S_8^a -\frac{1}{2}M_{S_8}^2 S_8^a S_8^a+y S_8^a\bar{q}_iT^a_{ij}q_j,
\end{aligned}  
\end{equation}
where the covariant derivative is $D_\mu^{ab}=\partial_\mu \delta^{ab} - g_s f^{abc} G^c_\mu$. Here, $q$ can be any of the standard model quarks, including both up-type and down-type.  However, considering the strong bound on the scalar octet's coupling to the light-quark flavors from the LHC dijet measurement~\cite{CMS:2019gwf}, we only consider the least-constrained scenario, which is a top-philic heavy resonance, where the only quark that the scalar octet couples to is the top quark.

\begin{figure}
  \centering
  \subfigure[] {\includegraphics[width=0.25\linewidth]{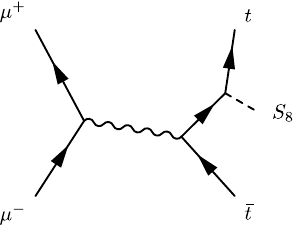}\label{fig:mumu2S8}}
  \qquad\qquad
  \subfigure[] {\includegraphics[width=0.25\linewidth]{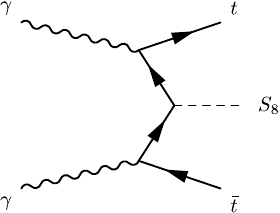}\label{fig:aa2S8}}
  \qquad\qquad
  \subfigure[] {\includegraphics[width=0.25\linewidth]{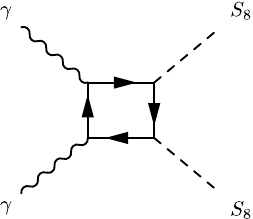}\label{fig:aabox}}
  \caption{Feynman diagrams for scalar octet $S_8$ in single production in (a) associated $t\bar{t}$ production and (b) photon-photon fusion, and(c) pair production induced at loop level.}
  \label{fig:S8}
\end{figure}

From the couplings of $S_8$ to tops and gluons, as seen in Eq.~\eqref{eq:S8}, pair production proceeds both through quark-antiquark annihilation, shown in Fig.~\ref{fig:qqbQCD}, and through gluon-gluon fusion, shown in Fig.~\ref{fig:gg2LGLG}.  Single production of $S_8$ occurs through top-antitop annihilation.  We do not take the top-quark PDF as our nominal choice, because of the large corrections from the large top mass in the collinear factorization. Instead, we compute top-antitop annihilation with explicit splittings, shown in Fig.~\ref{fig:mumu2S8}.  In this framework, the explicit diagrams are: final-state radiation in the annihilation shown in Fig.~\ref{fig:mumu2S8}, photon-photon fusion shown in Fig.~\ref{fig:aa2S8}, and
the box loop-induced photon-photon fusion shown in Fig.~\ref{fig:aabox}.

\begin{figure} 
  \centering
  \subfigure[] {\includegraphics[width=0.45\textwidth]{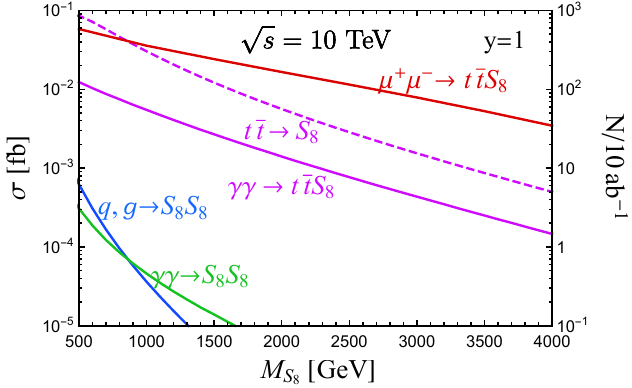}\label{fig:mumuSS_MS}}
  \qquad
  \subfigure[] {\includegraphics[width=0.45\textwidth]{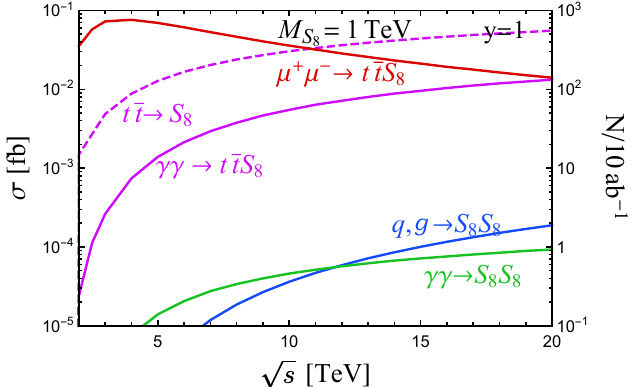}\label{fig:mumuSS_sqrts}}
  \caption{The cross section for the pair production $S_8 S_8$ and single production $S_8$ of a color-octet scalar $S_8$ as a function of mass (left) and as a function of center-of-mass energy (right).}
\end{figure}

The cross section is shown as a function of the scalar octet mass, $M_{S_8}$, at a $\sqrt{s}=10~\TeV$ muon collider in Fig.~\ref{fig:mumuSS_MS} and as a function of the collider energy in Fig.~\ref{fig:mumuSS_sqrts} for a $M_{S_8}=1~{\rm TeV}$ particle, where the benchmark coupling value is $y=1$.

For the top-antitop annihilation, if one uses the 
top-PDF approach,  we find it overestimates the cross section by a factor of $4 - 7$ with respect to the $\gamma\gamma\to t\bar{t}S_8$ approach. Two factors contribute to such a large correction.  First, the phase space is enlarged in collinear splitting $\gamma\to t\bar{t}$ into the top-quark PDF, which is only valid at high energies when top-quark mass is negligible, $Q\gg M_t$.  Second, additional contributions from the splitting $W/Z\to t\bar{q}$ (for $q=b,t$) is missing in the $\gamma\gamma$ fusion approach. However, exactly due to the same reason, the collinear splittings of $W/Z$ also suffer large threshold corrections due to the omission of the $W/Z$ masses. 
To be conservative, we mainly rely on the $\gamma\gamma$ fusion and the $\mu\mu$ annihilation for the projections in this work.
In comparison, the gluon-gluon fusion and box-induced photon-photon fusion are suppressed by one or two orders of magnitude.

\begin{figure} 
  \centering
  \includegraphics[width=0.55\textwidth]{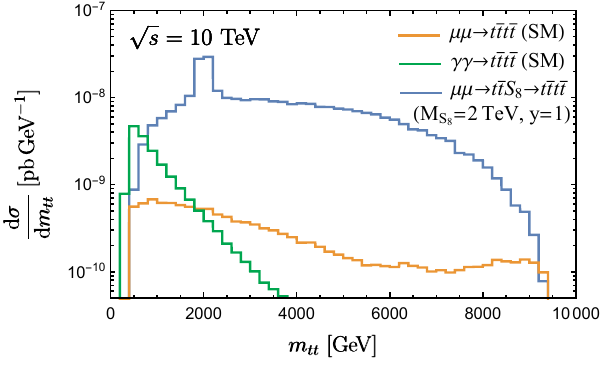}
  \caption{Distribution of $m_{tt}$ for a color-octet scalar $S_8$ at $\sqrt{s} = 10~{\rm TeV}$.  The signal point $(M_{S_8}, y) = (2~\TeV, y=1)$ is used (dark blue).  All other lines show various backgrounds. }
  \label{fig:S8bkg}
\end{figure}

\begin{figure} 
  \centering
  \subfigure[]{\includegraphics[width=0.45\textwidth]{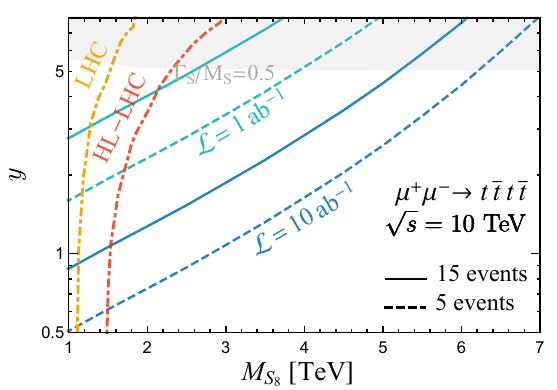}\label{fig:S8_reach}}
  \qquad
  \subfigure[]{\includegraphics[width=0.45\textwidth]{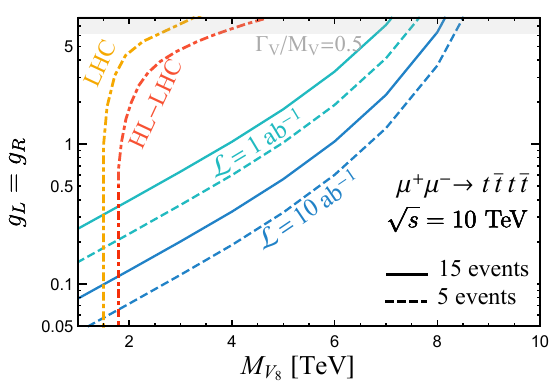}\label{fig:V8_reach}} 
  \caption{The discovery and exclusion limits of color-octet $S_8$ (left) and $V_8$ (right) in comparison with LHC and HL-LHC~\cite{Darme:2021gtt}. The shaded region indicates $\Gamma_{S}/M_{S}>0.5$ and $\Gamma_{V}/M_{V}>0.5$ where the theory prediction is unreliable. }
\end{figure}

Once the top-philic scalar octet is produced, it decays into a top-quark pair, which will end up with a final state of four tops since the top-philic $S_8$ is always produced in association with $t\bar{t}$.  The Standard Model background comes from four-top production, but without any diagrams with internal $S_8$ lines.  The corresponding cross sections are
\begin{equation}
\sigma(\mu^+\mu^-\to t\bar{t} t\bar{t})=3.5\times 10^{-3}~\textrm{fb},
\qquad\qquad
\sigma(\gamma\gamma\to t\bar{t} t\bar{t})=3.8\times 10^{-3}~\textrm{fb}.
\end{equation}
By reconstructing the top-pair invariant mass $m_{tt}$, we obtain the distributions for the signal and background, shown in Fig.~\ref{fig:S8bkg}.   We require the fiducial cuts on the final-state top quarks of
\begin{equation}\label{eq:fidS8}
|\eta_t| > 2.5, 
\qquad\qquad
\Delta R(t,t)>0.4.
\end{equation}
These cuts have a signal efficiency of 94\%.
In order to fully take advantage of all possible events, we consider both the leptonic and the hadronic channels of the top quark decays.
When the tops decay hadronically, it is very difficult to distinguish a top from an anti-top quark. Therefore, we include all 6 possible pairings of two tops in Fig.~\ref{fig:S8bkg}.  By applying an invariant mass cut
\begin{equation}
0.9 M_{S_8} < m_{tt} < 1.1M_{S_8},   
\end{equation}
the backgrounds are suppressed to a negligible level.
We apply a reconstruction efficiency of 50\% to each top quark~\cite{Kaplan:2008ie}.  Our projected discovery and exclusion sensitivity, based on the 5 and 15 events criterion, is shown in Fig.~\ref{fig:S8_reach}.
For scalar octets with masses between 1 and 2 TeV, a 10 TeV muon collider has comparable sensitivity to the LHC and the HL-LHC~\cite{Darme:2021gtt}, while at masses above 3 TeV a muon collider retains sensitivity whereas the LHC does not.  This is mainly driven by the higher partonic center-of-mass energy.

\subsection{Color-Octet Vector Bosons: Colorons}

A color-octet vector boson, often referred to as a ``coloron,'' with the representation
\begin{equation*}
\text{vector octet }(V_8): \quad\quad (8, 1)_{0},
\end{equation*}
is predicted in a variety of models, including as an axigluon in $SU(3) \times SU(3)$ models~\cite{Frampton:1987dn,Bagger:1987fz}, as a resonance in technicolor~\cite{Chivukula:1995dt}, or as a Kaluza-Klein (KK) gluon in extra-dimensional models~\cite{Dicus:2000hm}.  Similarly to the scalar case, we formulate a simplified Lagrangian for the color-octet vector as~\cite{Chivukula:2013xla}
\begin{equation}
\begin{aligned}
\mathcal{L}= -\frac{1}{4}F^a_{8,\mu\nu} F_8^{a,\mu\nu}+\frac{1}{2}M^2_V V_{8,\mu}^a V_8^{\mu,a}
 +  V_{8,\mu}^a \bar{q}_i \gamma^\mu (g_LP_L+g_R P_R)T^a_{ij} q_j,
\end{aligned}    
\end{equation}
where $F^a_{8,\mu\nu}=D_\mu V_{8,\nu}^a - D_\nu V_{8,\mu}^a$ and $D_\mu^{ab}=\partial_\mu \delta^{ab} - g_s f^{abc} G^c_\mu$.  As with the scalar octet, in this work, we consider the top-philic scenario, $q=t$, in order to avoid the strong exclusion limits from couplings to light-flavor quarks.

\begin{figure} 
  \centering
  \subfigure[] {\includegraphics[width=0.45\textwidth]{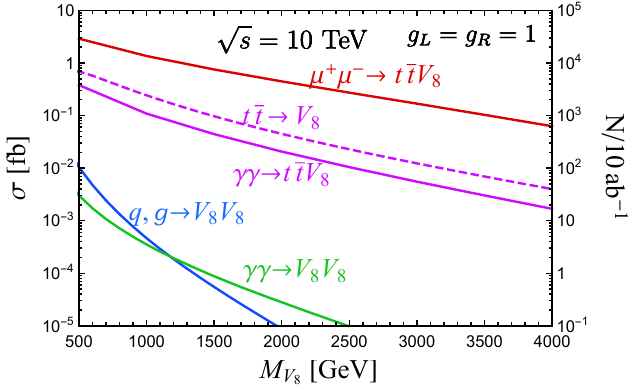}\label{fig:mumuVV_MV}}
  \qquad
  \subfigure[] {\includegraphics[width=0.45\textwidth]{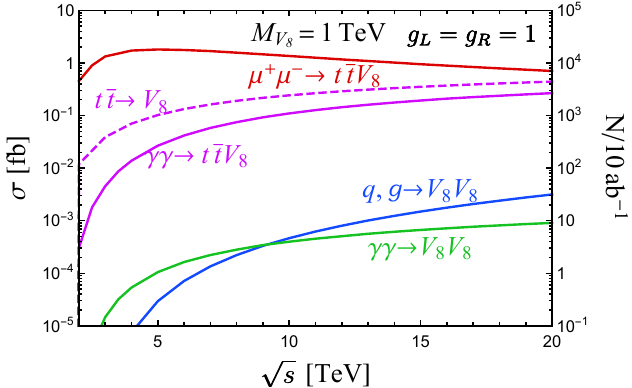}\label{fig:mumuVV_sqrts}}
  \caption{The cross section for the pair production $V_8 V_8$ and single production $V_8$ of a color-octet scalar $V_8$ as a function of mass (left) and as a function of center-of-mass energy (right).}
\end{figure}

The production mechanisms of color-octet vector bosons at a muon collider are similar to the scalar case, including production via muon annihilation in Fig.~\ref{fig:mumu2S8}, via photon-photon fusion in Fig.~\ref{fig:aa2S8}, via quark-antiquark annihilation in Fig.~\ref{fig:qqbQCD}, via gluon-gluon fusion in Fig.~\ref{fig:gg2LGLG}, and via box-induced photon fusion in Fig.~\ref{fig:aabox}.  The corresponding cross sections are presented in Fig.~\ref{fig:mumuVV_MV} and Fig.~\ref{fig:mumuVV_sqrts}.
The behavior is similar to the scalar octet with the only minor quantitative difference being that the top-antitop annihilation through the top-PDF approach only a factor of 2 overestimation with respect to the photon-photon fusion one.
As before, we use the $\gamma\gamma$ fusion cross approach instead of the top-quark PDF approach for projections, in addition to the $\mu\mu$ annihilation.

In our simulation, the signal distribution of $m_{tt}$ is similar to the signal distribution in Fig.~\ref{fig:S8bkg} and the backgrounds are identical.  Therefore, we apply the same cuts as Eq.~\eqref{eq:fidS8}, which gives an acceptance of about 90\% for the signal, and renders the background negligible. With the criterion of 5 or 15 signal events, we show the projected sensitivity for the color-octet vector in Fig.~\ref{fig:V8_reach}.
The sensitivity is slightly different from the scalar case because the state multiplicity for a scalar is 1 while for a vector it is 3.

\subsection{Color-Octet Fermions: Gluinos} 
\label{sec:gluino_pair}

In this last subsection, we consider a color-octet fermion, which corresponds to the gluino in supersymmetric theories.  The representation is
\begin{equation*}
\text{gluino}~(\tilde{g}): 
\qquad
(8, 1)_{0},
\end{equation*}
with the Lagrangian
\begin{align}
\mathcal{L}_{\tilde{g}}
= \frac{1}{2} \tilde{g} (i\slashed{D} - m_{\tilde{g}}) \tilde{g} 
= \frac{1}{2} (i\slashed{\partial} - m_{\tilde{g}}) \tilde{g}
+\frac{1}{2} i g_s f^{abc} \tilde{g}^a \gamma^\mu \tilde{g}^b G^c_\mu. 
\end{align}
The interaction with squarks is also included and takes the form $\sqrt{2} g_s \Tilde{q}^* T^a \tilde{g}^a q$.  We include all generations and both chirality partners of squarks.
%

\begin{figure}
  \centering
  \subfigure[]{\includegraphics[width=0.24\linewidth]{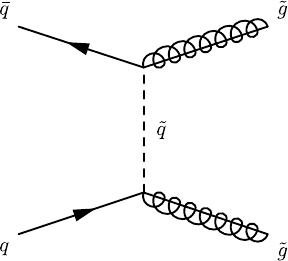}\label{fig:qqb2gogo}}
  \qquad\qquad
  \subfigure[]{\includegraphics[width=0.24\linewidth]{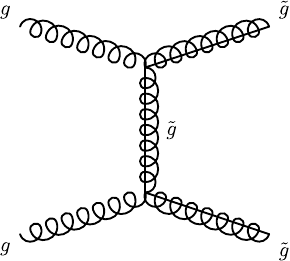}\label{fig:gg2gogo}}
  \caption{Feynman diagrams for the pair production of gluinos mediated by (a) a $t$-channel squark and (b) a $t$-channel gluino.}
  \label{eq:feynman_gluino}
\end{figure}

At a muon collider, gluinos can be produced in pairs through quark or gluon scattering.   For quark-antiquark scattering, we have the $s$-channel diagram shown in Fig.~\ref{fig:qqbQCD}, and the $t$-channel squark diagrams shown in Fig.~\ref{fig:qqb2gogo}. For the gluon-gluon fusion, we have an $s$-channel diagram mediated by a gluon like in Fig.~\ref{fig:gg2muLG} and a $t$-channel gluino diagram shown in Fig.~\ref{fig:gg2gogo}.  In addition, there is photon-photon fusion mediated through the quark and squark loops, like the diagram shown in Fig.~\ref{fig:aabox}.

\begin{figure}
  \centering
  \includegraphics[width=0.45\textwidth]{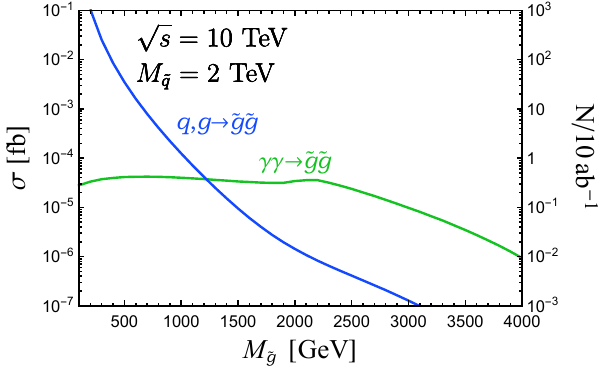}
  \qquad
    \includegraphics[width=0.45\textwidth]{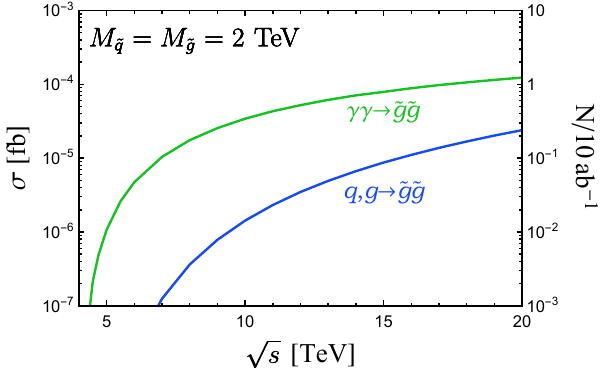}
  \caption{The cross section for the pair production $\tilde{g} \tilde{g}$ of a color-octet fermion $\tilde{g}$ as a function of mass (left) and as a function of center-of-mass energy (right).}
\label{fig:mumugogo}
\end{figure}

In Fig.~\ref{fig:mumugogo}, we present the cross section for gluino pair production at a high-energy muon collider.  For illustration, we take the squarks to be degenerate at a mass of $M_{\tilde{q}}=2~\TeV$.  The current LHC exclusion limit on gluinos is 2.2 TeV~\cite{ATLAS:2023afl} which corresponds to a cross section at a 10 TeV muon collider of less than $10^{-4}$~fb.  The expected number of gluino events is less than 1 with an integrated luminosity of $10~\textrm{ab}^{-1}$ and we will therefore not make a detailed reach estimate.

When the squark spectrum is not degenerate, the muon annihilation cross section is very sensitive to the splittings between the masses of the squarks~\cite{Kileng:1994kw,Kileng:1994vc}.  A mass splitting of 10\% can lead to a muon annihilation cross section enhancement of up to 3 orders of magnitude.  Despite this large increase, over much of the parameter range, the photon fusion contribution to the cross section is still dominant, leading to an unaffected total cross section.  This sensitivity is present in muon annihilation because the loop part of the gluino current vanishes for photon propagators.  For $Z$ propagators the up-type and down-type quarks interfere destructively leading to a small cross section (see section 4.1 Ref.~\cite{Berge:2003bz} for further discussion).

\begin{figure}
  \centering
  \includegraphics[width=0.45\textwidth]{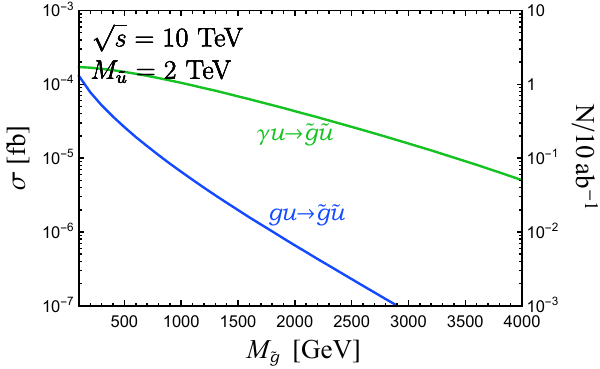}
  \qquad
  \includegraphics[width=0.45\textwidth]{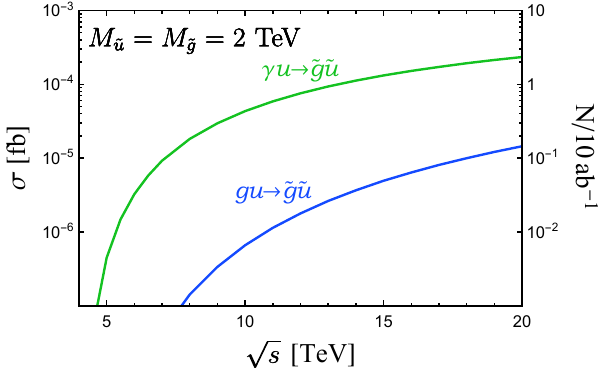}
  \caption{The cross section for the associated production $\tilde{g} \tilde{u}$ of a gluino $\tilde{g}$ and an up-type squark $\tilde{u}$ as a function of mass (left) and as a function of center-of-mass energy (right).}
  \label{fig:mumugq}
\end{figure}

When squarks are present in the spectrum, gluinos can be also produced in association with a squark. In Fig.~\ref{fig:mumugq}, we compute the associated production cross section of a gluino and an up-type squark.  The squark mass is fixed at $M_{\tilde{u}}=2~\TeV$.  The main production mechanism comes from a photon scattering with an up-type quark.  Again, with the current LHC limits on the gluino of $M_{\tilde{g}}>2.2~\TeV$ and on squarks of $M_{\tilde{q}}>1.7~\TeV$~\cite{ATLAS:2023afl}, we expect the corresponding number of events to be less than 1 at a 10 TeV muon collider.

\section{Conclusions}
\label{sec:conclusion}

In this work, we have explored the phenomenology of colored particles at future high-energy muon colliders.  For a wide range of new particles, varying by their spins and quantum numbers, we have presented both a detailed breakdown of their pair and single production rates across all relevant channels and an estimated reach with a 10 TeV muon collider.

Many colored particles, by nature of their roles in BSM theories, have predicted electroweak representations and hypercharges.  When the electroweak charge is non-zero, the channel of $\mu^+\mu^-$ annihilation via $\gamma^*/Z^*$ exchange is always open once above the pair production threshold, leading to sizable production rates and competitive projected reach.  This is the situation for color-triplet fermions $(T)$ and color-triplet scalars $(\tilde q)$ where the pair production rate is roughly 2 fb and 0.5 fb, respectively.  With $10~{\rm ab}^{-1}$ of data, a muon collider can discover these particles close to 5 TeV which is better than both the LHC and HL-LHC.  For single production these limits extend past 5 TeV, depending on the model-dependent coupling, and also are substantially stronger than the HL-LHC.

The diquark $(D_6)$ cross section at a muon collider is of a similar size, but the potential reach pales compared to the LHC.  The LHC has a large rate for single diquark production whereas single diquark production at a muon collider requires a quark parton from the muon PDF and a quark parton from the anti-muon PDF, resulting in a large suppression in production rate, as compared to hadron colliders.

Next, we studied scalar leptoquarks $(S_{\ell_q})$ and fermion leptogluons $(\ell_g)$ which can be produced in pairs in $\mu^+\mu^-$ annihilation or can be singly produced from an incoming muon and an incoming quark parton.  For leptoquark couplings larger than 0.05 single production leads to better sensitivity.  At the LHC leptoquark single production proceeds in an analogous manner.  We show that the leptoquark reach, assuming a coupling to muons, is much stronger than what can be achieved at the LHC.  The conclusion is the same for leptogluons.

Finally, we analyzed color-octet particles in the form of a color-octet scalar $(S_8)$, color-octet vector $(V_8)$, and color-octet fermion, gluino ($\tilde g$). For the scalar and the vector, we consider the top-philic case where the only quark that couples to the new particle is the top quark.  Consequently, in a muon collider, both single production and pair production lead to a final state of $t\bar{t}t\bar{t}$, which varies by resonance structure.  As the scalar or vector is produced from fusion $t\bar{t}$ or radiated from a $t$ or $\bar{t}$ the cross sections are still sizable with rates around 0.05 fb for scalars and 1 fb for vectors.  This leads to strong constraints that exceed those of the LHC and the HL-LHC.  The color-octet fermion, better known as the gluino, is a bit different since it either needs to be pair produced or produced in association with a squark.  The final state in both cases contains two heavy particles and the initial state requires colored particles.  Consequently, the rate is below $10^{-4}$ fb. Although a handful of events could be collected, there is no opportunity to exceed the LHC limits. 

In summary, despite the fact that a muon collider collides two leptons with no QCD charges, such a collider still has considerable sensitivity to study heavy colored particles at high energies.  We illustrated this directly through rate calculations and simulated sensitivities.  Color-triplet fermions and scalars are exemplars as these are common in BSM theories and their limits benefit immensely from a high-energy muon collider.  Leptoquarks and leptogluons are also noteworthy cases where a muon collider drives their mass limits at high masses.  Such calculations, in general, require a proper inclusion of colored particles into the PDF of the muon which we do in this work including the full parton evolution.  While we survey several of the most common new physics possibilities, we expect this work to serve as a starting point for future studies of new physics at muon colliders.  Full signal and background simulations would be ideal follow-up work as would studies for particles with different quantum numbers, as well as the collider and detector performance and requirements.

As a final remark, although our studies have been fully focused on a muon collider, our results are largely applicable to high-energy $e^+e^-$ collisions as well. For the dominant pair production mechanism, the annihilation of $\mu^+\mu^-$ and $e^+e^-$ would yield essentially the same outcome. The difference comes when involving the initial-state photon processes, which would be enhanced in $e^+e^-$ collisions, due to an increase in the collinear photon splitting by a factor $\ln(Q/m_e)/\ln(Q/m_\mu)$, which is $\approx 1.6$ at the TeV scale. 
The only dramatic difference would be for flavor-dependent physics, which distinguishes a muon from an electron. In such a scenario, the two colliders would be complementary to probe the details of the underlying lepton-flavor dynamics. 

\begin{acknowledgements}
The authors thank Si Wang for participation at the early stage of this project. This work was supported in part by the US Department of Energy under grant No. DE-SC0007914 and in part by Pitt PACC. TH would like to thank the Aspen Center for Physics, where part of this work is complete, which is supported by the National Science Foundation grant No. PHY-1607611, and the hospitality extended to him during the final stage of the project by CERN TH Department and the Kavli Institute for Theoretical Physics (KITP), which was supported by grant NSF PHY-2309135. ML and KX are also supported by the National Science Foundation under grant No. PHY-2112829, PHY-2310291 and PHY-2310497.  We used high-performance computing resources from the Pitt CRC, MSU HPCC, and SMU M3.
\end{acknowledgements}

\bibliographystyle{utphys}
\bibliography{refs}

\providecommand{\href}[2]{#2}\begingroup\raggedright\begin{thebibliography}{100}

\bibitem{ATLAS:2012yve}
{\bfseries ATLAS} Collaboration, G.~Aad {\em et~al.}, ``{Observation of a new particle in the search for the Standard Model Higgs boson with the ATLAS detector at the LHC},'' \href{http://dx.doi.org/10.1016/j.physletb.2012.08.020}{{\em Phys. Lett. B} {\bfseries 716} (2012) 1--29}, \href{http://arxiv.org/abs/1207.7214}{{\ttfamily arXiv:1207.7214 [hep-ex]}}.

\bibitem{CMS:2012qbp}
{\bfseries CMS} Collaboration, S.~Chatrchyan {\em et~al.}, ``{Observation of a New Boson at a Mass of 125 GeV with the CMS Experiment at the LHC},'' \href{http://dx.doi.org/10.1016/j.physletb.2012.08.021}{{\em Phys. Lett. B} {\bfseries 716} (2012) 30--61}, \href{http://arxiv.org/abs/1207.7235}{{\ttfamily arXiv:1207.7235 [hep-ex]}}.

\bibitem{P5:2023wyd}
{\bfseries P5} Collaboration, S.~Asai {\em et~al.}, ``{Exploring the Quantum Universe: Pathways to Innovation and Discovery in Particle Physics},'' \href{http://arxiv.org/abs/2407.19176}{{\ttfamily arXiv:2407.19176 [hep-ex]}}.

\bibitem{Delahaye:2019omf}
J.~P. Delahaye, M.~Diemoz, K.~Long, B.~Mansouli\'e, N.~Pastrone, L.~Rivkin, D.~Schulte, A.~Skrinsky, and A.~Wulzer, ``{Muon Colliders},'' \href{http://arxiv.org/abs/1901.06150}{{\ttfamily arXiv:1901.06150 [physics.acc-ph]}}.

\bibitem{Bartosik:2020xwr}
N.~Bartosik {\em et~al.}, ``{Detector and Physics Performance at a Muon Collider},'' \href{http://dx.doi.org/10.1088/1748-0221/15/05/P05001}{{\em JINST} {\bfseries 15} no.~05, (2020) P05001}, \href{http://arxiv.org/abs/2001.04431}{{\ttfamily arXiv:2001.04431 [hep-ex]}}.

\bibitem{Schulte:2021hgo}
D.~Schulte, J.-P. Delahaye, M.~Diemoz, K.~Long, B.~Mansouli\'e, N.~Pastrone, L.~Rivkin, A.~Skrinsky, and A.~Wulzer, ``{Prospects on Muon Colliders},'' \href{http://dx.doi.org/10.22323/1.390.0703}{{\em PoS} {\bfseries ICHEP2020} (2021) 703}.

\bibitem{Long:2020wfp}
K.~Long, D.~Lucchesi, M.~Palmer, N.~Pastrone, D.~Schulte, and V.~Shiltsev, ``{Muon colliders to expand frontiers of particle physics},'' \href{http://dx.doi.org/10.1038/s41567-020-01130-x}{{\em Nature Phys.} {\bfseries 17} no.~3, (2021) 289--292}, \href{http://arxiv.org/abs/2007.15684}{{\ttfamily arXiv:2007.15684 [physics.acc-ph]}}.

\bibitem{MuonCollider:2022xlm}
{\bfseries Muon Collider} Collaboration, J.~de~Blas {\em et~al.}, ``{The physics case of a 3 TeV muon collider stage},'' \href{http://arxiv.org/abs/2203.07261}{{\ttfamily arXiv:2203.07261 [hep-ph]}}.

\bibitem{Aime:2022flm}
C.~Aime {\em et~al.}, ``{Muon Collider Physics Summary},'' \href{http://arxiv.org/abs/2203.07256}{{\ttfamily arXiv:2203.07256 [hep-ph]}}.

\bibitem{Black:2022cth}
K.~M. Black {\em et~al.}, ``{Muon Collider Forum report},'' \href{http://dx.doi.org/10.1088/1748-0221/19/02/T02015}{{\em JINST} {\bfseries 19} no.~02, (2024) T02015}, \href{http://arxiv.org/abs/2209.01318}{{\ttfamily arXiv:2209.01318 [hep-ex]}}.

\bibitem{Accettura:2023ked}
C.~Accettura {\em et~al.}, ``{Towards a muon collider},'' \href{http://dx.doi.org/10.1140/epjc/s10052-023-11889-x}{{\em Eur. Phys. J. C} {\bfseries 83} no.~9, (2023) 864}, \href{http://arxiv.org/abs/2303.08533}{{\ttfamily arXiv:2303.08533 [physics.acc-ph]}}. [Erratum: Eur.Phys.J.C 84, 36 (2024)].

\bibitem{Ciafaloni:2000df}
M.~Ciafaloni, P.~Ciafaloni, and D.~Comelli, ``{Bloch-Nordsieck violating electroweak corrections to inclusive TeV scale hard processes},'' \href{http://dx.doi.org/10.1103/PhysRevLett.84.4810}{{\em Phys. Rev. Lett.} {\bfseries 84} (2000) 4810--4813}, \href{http://arxiv.org/abs/hep-ph/0001142}{{\ttfamily arXiv:hep-ph/0001142}}.

\bibitem{Ciafaloni:2001mu}
M.~Ciafaloni, P.~Ciafaloni, and D.~Comelli, ``{Towards collinear evolution equations in electroweak theory},'' \href{http://dx.doi.org/10.1103/PhysRevLett.88.102001}{{\em Phys. Rev. Lett.} {\bfseries 88} (2002) 102001}, \href{http://arxiv.org/abs/hep-ph/0111109}{{\ttfamily arXiv:hep-ph/0111109}}.

\bibitem{Chen:2016wkt}
J.~Chen, T.~Han, and B.~Tweedie, ``{Electroweak Splitting Functions and High Energy Showering},'' \href{http://dx.doi.org/10.1007/JHEP11(2017)093}{{\em JHEP} {\bfseries 11} (2017) 093}, \href{http://arxiv.org/abs/1611.00788}{{\ttfamily arXiv:1611.00788 [hep-ph]}}.

\bibitem{Costantini:2020stv}
A.~Costantini, F.~De~Lillo, F.~Maltoni, L.~Mantani, O.~Mattelaer, R.~Ruiz, and X.~Zhao, ``{Vector boson fusion at multi-TeV muon colliders},'' \href{http://dx.doi.org/10.1007/JHEP09(2020)080}{{\em JHEP} {\bfseries 09} (2020) 080}, \href{http://arxiv.org/abs/2005.10289}{{\ttfamily arXiv:2005.10289 [hep-ph]}}.

\bibitem{BuarqueFranzosi:2021wrv}
D.~Buarque~Franzosi {\em et~al.}, ``{Vector boson scattering processes: Status and prospects},'' \href{http://dx.doi.org/10.1016/j.revip.2022.100071}{{\em Rev. Phys.} {\bfseries 8} (2022) 100071}, \href{http://arxiv.org/abs/2106.01393}{{\ttfamily arXiv:2106.01393 [hep-ph]}}.

\bibitem{Han:2020uid}
T.~Han, Y.~Ma, and K.~Xie, ``{High energy leptonic collisions and electroweak parton distribution functions},'' \href{http://dx.doi.org/10.1103/PhysRevD.103.L031301}{{\em Phys. Rev. D} {\bfseries 103} no.~3, (2021) L031301}, \href{http://arxiv.org/abs/2007.14300}{{\ttfamily arXiv:2007.14300 [hep-ph]}}.

\bibitem{AlAli:2021let}
H.~Al~Ali {\em et~al.}, ``{The muon Smasher\textquoteright{}s guide},'' \href{http://dx.doi.org/10.1088/1361-6633/ac6678}{{\em Rept. Prog. Phys.} {\bfseries 85} no.~8, (2022) 084201}, \href{http://arxiv.org/abs/2103.14043}{{\ttfamily arXiv:2103.14043 [hep-ph]}}.

\bibitem{Bauer:2017isx}
C.~W. Bauer, N.~Ferland, and B.~R. Webber, ``{Standard Model Parton Distributions at Very High Energies},'' \href{http://dx.doi.org/10.1007/JHEP08(2017)036}{{\em JHEP} {\bfseries 08} (2017) 036}, \href{http://arxiv.org/abs/1703.08562}{{\ttfamily arXiv:1703.08562 [hep-ph]}}.

\bibitem{Fornal:2018znf}
B.~Fornal, A.~V. Manohar, and W.~J. Waalewijn, ``{Electroweak Gauge Boson Parton Distribution Functions},'' \href{http://dx.doi.org/10.1007/JHEP05(2018)106}{{\em JHEP} {\bfseries 05} (2018) 106}, \href{http://arxiv.org/abs/1803.06347}{{\ttfamily arXiv:1803.06347 [hep-ph]}}.

\bibitem{Bauer:2018arx}
C.~W. Bauer and B.~R. Webber, ``{Polarization Effects in Standard Model Parton Distributions at Very High Energies},'' \href{http://dx.doi.org/10.1007/JHEP03(2019)013}{{\em JHEP} {\bfseries 03} (2019) 013}, \href{http://arxiv.org/abs/1808.08831}{{\ttfamily arXiv:1808.08831 [hep-ph]}}.

\bibitem{Han:2021kes}
T.~Han, Y.~Ma, and K.~Xie, ``{Quark and gluon contents of a lepton at high energies},'' \href{http://dx.doi.org/10.1007/JHEP02(2022)154}{{\em JHEP} {\bfseries 02} (2022) 154}, \href{http://arxiv.org/abs/2103.09844}{{\ttfamily arXiv:2103.09844 [hep-ph]}}.

\bibitem{Garosi:2023bvq}
F.~Garosi, D.~Marzocca, and S.~Trifinopoulos, ``{LePDF: Standard Model PDFs for high-energy lepton colliders},'' \href{http://dx.doi.org/10.1007/JHEP09(2023)107}{{\em JHEP} {\bfseries 09} (2023) 107}, \href{http://arxiv.org/abs/2303.16964}{{\ttfamily arXiv:2303.16964 [hep-ph]}}.

\bibitem{ATLAS:2020syg}
{\bfseries ATLAS} Collaboration, G.~Aad {\em et~al.}, ``{Search for squarks and gluinos in final states with jets and missing transverse momentum using 139 fb$^{-1}$ of $\sqrt{s}$ =13 TeV $pp$ collision data with the ATLAS detector},'' \href{http://dx.doi.org/10.1007/JHEP02(2021)143}{{\em JHEP} {\bfseries 02} (2021) 143}, \href{http://arxiv.org/abs/2010.14293}{{\ttfamily arXiv:2010.14293 [hep-ex]}}.

\bibitem{ATLAS:2023afl}
{\bfseries ATLAS} Collaboration, G.~Aad {\em et~al.}, ``{Search for pair production of squarks or gluinos decaying via sleptons or weak bosons in final states with two same-sign or three leptons with the ATLAS detector},'' \href{http://dx.doi.org/10.1007/JHEP02(2024)107}{{\em JHEP} {\bfseries 02} (2024) 107}, \href{http://arxiv.org/abs/2307.01094}{{\ttfamily arXiv:2307.01094 [hep-ex]}}.

\bibitem{CMS:2023ktc}
{\bfseries CMS} Collaboration, A.~Tumasyan {\em et~al.}, ``{Search for top squarks in the four-body decay mode with single lepton final states in proton-proton collisions at $ \sqrt{s} $ = 13 TeV},'' \href{http://dx.doi.org/10.1007/JHEP06(2023)060}{{\em JHEP} {\bfseries 06} (2023) 060}, \href{http://arxiv.org/abs/2301.08096}{{\ttfamily arXiv:2301.08096 [hep-ex]}}.

\bibitem{ATLAS:2024rcx}
{\bfseries ATLAS} Collaboration, G.~Aad {\em et~al.}, ``{Search for new phenomena with top-quark pairs and large missing transverse momentum using $140$ fb$^{-1}$ of pp collision data at $ \sqrt{s} = $ $13$ TeV with the ATLAS detector},'' \href{http://dx.doi.org/10.1007/JHEP03(2024)139}{{\em JHEP} {\bfseries 03} (2024) 139}, \href{http://arxiv.org/abs/2401.13430}{{\ttfamily arXiv:2401.13430 [hep-ex]}}.

\bibitem{ParticleDataGroup:2024cfk}
{\bfseries Particle Data Group} Collaboration, S.~Navas {\em et~al.}, ``{Review of particle physics},'' \href{http://dx.doi.org/10.1103/PhysRevD.110.030001}{{\em Phys. Rev. D} {\bfseries 110} no.~3, (2024) 030001}.

\bibitem{Banerjee:2024zvg}
A.~Banerjee, E.~Bergeaas~Kuutmann, V.~Ellajosyula, R.~Enberg, G.~Ferretti, and L.~Panizzi, ``{Vector-like quarks: Status and new directions at the LHC},'' \href{http://dx.doi.org/10.21468/SciPostPhysCore.7.4.079}{{\em SciPost Phys. Core} {\bfseries 7} (2024) 079}, \href{http://arxiv.org/abs/2406.09193}{{\ttfamily arXiv:2406.09193 [hep-ph]}}.

\bibitem{ATLAS:2018zrp}
{\bfseries ATLAS} Collaboration, ``{ATLAS sensitivity to top squark pair production at the HL-LHC},''.

\bibitem{ATLAS:2019mfr}
{\bfseries ATLAS, CMS} Collaboration, ``{Addendum to the report on the physics at the HL-LHC, and perspectives for the HE-LHC: Collection of notes from ATLAS and CMS},'' \href{http://dx.doi.org/10.23731/CYRM-2019-007.Addendum}{{\em CERN Yellow Rep. Monogr.} {\bfseries 7} (2019) Addendum}, \href{http://arxiv.org/abs/1902.10229}{{\ttfamily arXiv:1902.10229 [hep-ex]}}.

\bibitem{Baer:2023uwo}
H.~Baer, V.~Barger, J.~Dutta, D.~Sengupta, and K.~Zhang, ``{Top squarks from the landscape at high luminosity LHC},'' \href{http://dx.doi.org/10.1103/PhysRevD.108.075027}{{\em Phys. Rev. D} {\bfseries 108} no.~7, (2023) 075027}, \href{http://arxiv.org/abs/2307.08067}{{\ttfamily arXiv:2307.08067 [hep-ph]}}.

\bibitem{Berger:1996un}
M.~S. Berger and W.~Merritt, ``{Discovering new particles at colliders},'' {\em eConf} {\bfseries C960625} (1996) NEW144, \href{http://arxiv.org/abs/hep-ph/9611386}{{\ttfamily arXiv:hep-ph/9611386}}.

\bibitem{Asadi:2021gah}
P.~Asadi, R.~Capdevilla, C.~Cesarotti, and S.~Homiller, ``{Searching for leptoquarks at future muon colliders},'' \href{http://dx.doi.org/10.1007/JHEP10(2021)182}{{\em JHEP} {\bfseries 10} (2021) 182}, \href{http://arxiv.org/abs/2104.05720}{{\ttfamily arXiv:2104.05720 [hep-ph]}}.

\bibitem{Huang:2021biu}
G.-y. Huang, S.~Jana, F.~S. Queiroz, and W.~Rodejohann, ``{Probing the RK(*) anomaly at a muon collider},'' \href{http://dx.doi.org/10.1103/PhysRevD.105.015013}{{\em Phys. Rev. D} {\bfseries 105} no.~1, (2022) 015013}, \href{http://arxiv.org/abs/2103.01617}{{\ttfamily arXiv:2103.01617 [hep-ph]}}.

\bibitem{Bandyopadhyay:2021pld}
P.~Bandyopadhyay, A.~Karan, R.~Mandal, and S.~Parashar, ``{Distinguishing signatures of scalar leptoquarks at hadron and muon colliders},'' \href{http://dx.doi.org/10.1140/epjc/s10052-022-10809-9}{{\em Eur. Phys. J. C} {\bfseries 82} no.~10, (2022) 916}, \href{http://arxiv.org/abs/2108.06506}{{\ttfamily arXiv:2108.06506 [hep-ph]}}.

\bibitem{Qian:2021ihf}
S.~Qian, C.~Li, Q.~Li, F.~Meng, J.~Xiao, T.~Yang, M.~Lu, and Z.~You, ``{Searching for heavy leptoquarks at a muon collider},'' \href{http://dx.doi.org/10.1007/JHEP12(2021)047}{{\em JHEP} {\bfseries 12} (2021) 047}, \href{http://arxiv.org/abs/2109.01265}{{\ttfamily arXiv:2109.01265 [hep-ph]}}.

\bibitem{Lv:2022pts}
G.-S. Lv, X.-M. Cui, Y.-Q. Li, and Y.-B. Liu, ``{Pair production of the vectorlike top partner at future muon collider},'' \href{http://dx.doi.org/10.1016/j.nuclphysb.2022.116016}{{\em Nucl. Phys. B} {\bfseries 985} (2022) 116016}.

\bibitem{Azatov:2022itm}
A.~Azatov, F.~Garosi, A.~Greljo, D.~Marzocca, J.~Salko, and S.~Trifinopoulos, ``{New physics in b \textrightarrow{} s\ensuremath{\mu}\ensuremath{\mu}: FCC-hh or a muon collider?},'' \href{http://dx.doi.org/10.1007/JHEP10(2022)149}{{\em JHEP} {\bfseries 10} (2022) 149}, \href{http://arxiv.org/abs/2205.13552}{{\ttfamily arXiv:2205.13552 [hep-ph]}}.

\bibitem{Belyaev:2023yym}
A.~Belyaev, R.~S. Chivukula, B.~Fuks, E.~H. Simmons, and X.~Wang, ``{Vectorlike top quark production via an electroweak dipole moment at a muon collider},'' \href{http://dx.doi.org/10.1103/PhysRevD.108.035016}{{\em Phys. Rev. D} {\bfseries 108} no.~3, (2023) 035016}, \href{http://arxiv.org/abs/2306.11097}{{\ttfamily arXiv:2306.11097 [hep-ph]}}.

\bibitem{Ghosh:2023xbj}
N.~Ghosh, S.~K. Rai, and T.~Samui, ``{Search for a leptoquark and vector-like lepton in a muon collider},'' \href{http://dx.doi.org/10.1016/j.nuclphysb.2024.116564}{{\em Nucl. Phys. B} {\bfseries 1004} (2024) 116564}, \href{http://arxiv.org/abs/2309.07583}{{\ttfamily arXiv:2309.07583 [hep-ph]}}.

\bibitem{Liu:2023jta}
D.~Liu, L.-T. Wang, and K.-P. Xie, ``{Composite resonances at a 10 TeV muon collider},'' \href{http://dx.doi.org/10.1007/JHEP04(2024)084}{{\em JHEP} {\bfseries 04} (2024) 084}, \href{http://arxiv.org/abs/2312.09117}{{\ttfamily arXiv:2312.09117 [hep-ph]}}.

\bibitem{Bhaskar:2024snl}
A.~Bhaskar and M.~Mitra, ``{Boosted top quark inspired leptoquark searches at the muon collider},'' \href{http://arxiv.org/abs/2409.15992}{{\ttfamily arXiv:2409.15992 [hep-ph]}}.

\bibitem{Nilles:1983ge}
H.~P. Nilles, ``{Supersymmetry, Supergravity and Particle Physics},'' \href{http://dx.doi.org/10.1016/0370-1573(84)90008-5}{{\em Phys. Rept.} {\bfseries 110} (1984) 1--162}.

\bibitem{Haber:1984rc}
H.~E. Haber and G.~L. Kane, ``{The Search for Supersymmetry: Probing Physics Beyond the Standard Model},'' \href{http://dx.doi.org/10.1016/0370-1573(85)90051-1}{{\em Phys. Rept.} {\bfseries 117} (1985) 75--263}.

\bibitem{Martin:1997ns}
S.~P. Martin, ``{A Supersymmetry primer},'' \href{http://dx.doi.org/10.1142/9789812839657_0001}{{\em Adv. Ser. Direct. High Energy Phys.} {\bfseries 18} (1998) 1--98}, \href{http://arxiv.org/abs/hep-ph/9709356}{{\ttfamily arXiv:hep-ph/9709356}}.

\bibitem{Buchmueller:2013exa}
O.~Buchmueller and J.~Marrouche, ``{Universal mass limits on gluino and third-generation squarks in the context of Natural-like SUSY spectra},'' \href{http://dx.doi.org/10.1142/S0217751X14500328}{{\em Int. J. Mod. Phys. A} {\bfseries 29} no.~06, (2014) 1450032}, \href{http://arxiv.org/abs/1304.2185}{{\ttfamily arXiv:1304.2185 [hep-ph]}}.

\bibitem{Papucci:2011wy}
M.~Papucci, J.~T. Ruderman, and A.~Weiler, ``{Natural SUSY Endures},'' \href{http://dx.doi.org/10.1007/JHEP09(2012)035}{{\em JHEP} {\bfseries 09} (2012) 035}, \href{http://arxiv.org/abs/1110.6926}{{\ttfamily arXiv:1110.6926 [hep-ph]}}.

\bibitem{Agashe:2004rs}
K.~Agashe, R.~Contino, and A.~Pomarol, ``{The Minimal composite Higgs model},'' \href{http://dx.doi.org/10.1016/j.nuclphysb.2005.04.035}{{\em Nucl. Phys. B} {\bfseries 719} (2005) 165--187}, \href{http://arxiv.org/abs/hep-ph/0412089}{{\ttfamily arXiv:hep-ph/0412089}}.

\bibitem{Schmaltz:2005ky}
M.~Schmaltz and D.~Tucker-Smith, ``{Little Higgs review},'' \href{http://dx.doi.org/10.1146/annurev.nucl.55.090704.151502}{{\em Ann. Rev. Nucl. Part. Sci.} {\bfseries 55} (2005) 229--270}, \href{http://arxiv.org/abs/hep-ph/0502182}{{\ttfamily arXiv:hep-ph/0502182}}.

\bibitem{Panico:2015jxa}
G.~Panico and A.~Wulzer, \href{http://dx.doi.org/10.1007/978-3-319-22617-0}{{\em {The Composite Nambu-Goldstone Higgs}}}, vol.~913.
\newblock Springer, 2016.
\newblock \href{http://arxiv.org/abs/1506.01961}{{\ttfamily arXiv:1506.01961 [hep-ph]}}.

\bibitem{Kilic:2009mi}
C.~Kilic, T.~Okui, and R.~Sundrum, ``{Vectorlike Confinement at the LHC},'' \href{http://dx.doi.org/10.1007/JHEP02(2010)018}{{\em JHEP} {\bfseries 02} (2010) 018}, \href{http://arxiv.org/abs/0906.0577}{{\ttfamily arXiv:0906.0577 [hep-ph]}}.

\bibitem{Arkani-Hamed:2016kpz}
N.~Arkani-Hamed, R.~T. D'Agnolo, M.~Low, and D.~Pinner, ``{Unification and New Particles at the LHC},'' \href{http://dx.doi.org/10.1007/JHEP11(2016)082}{{\em JHEP} {\bfseries 11} (2016) 082}, \href{http://arxiv.org/abs/1608.01675}{{\ttfamily arXiv:1608.01675 [hep-ph]}}.

\bibitem{Hewett:1988xc}
J.~L. Hewett and T.~G. Rizzo, ``{Low-Energy Phenomenology of Superstring Inspired E(6) Models},'' \href{http://dx.doi.org/10.1016/0370-1573(89)90071-9}{{\em Phys. Rept.} {\bfseries 183} (1989) 193}.

\bibitem{Plehn:2008ae}
T.~Plehn and T.~M.~P. Tait, ``{Seeking Sgluons},'' \href{http://dx.doi.org/10.1088/0954-3899/36/7/075001}{{\em J. Phys. G} {\bfseries 36} (2009) 075001}, \href{http://arxiv.org/abs/0810.3919}{{\ttfamily arXiv:0810.3919 [hep-ph]}}.

\bibitem{Frampton:2009rk}
P.~H. Frampton, J.~Shu, and K.~Wang, ``{Axigluon as Possible Explanation for p anti-p ---\ensuremath{>} t anti-t Forward-Backward Asymmetry},'' \href{http://dx.doi.org/10.1016/j.physletb.2009.12.043}{{\em Phys. Lett. B} {\bfseries 683} (2010) 294--297}, \href{http://arxiv.org/abs/0911.2955}{{\ttfamily arXiv:0911.2955 [hep-ph]}}.

\bibitem{Bhaskar:2023ftn}
A.~Bhaskar, A.~Das, T.~Mandal, S.~Mitra, and R.~Sharma, ``{Fresh look at the LHC limits on scalar leptoquarks},'' \href{http://dx.doi.org/10.1103/PhysRevD.109.055018}{{\em Phys. Rev. D} {\bfseries 109} no.~5, (2024) 055018}, \href{http://arxiv.org/abs/2312.09855}{{\ttfamily arXiv:2312.09855 [hep-ph]}}.

\bibitem{CMS:2024bnj}
{\bfseries CMS} Collaboration, A.~Hayrapetyan {\em et~al.}, ``{Search for pair production of scalar and vector leptoquarks decaying to muons and bottom quarks in proton-proton collisions at s=13\,\,TeV},'' \href{http://dx.doi.org/10.1103/PhysRevD.109.112003}{{\em Phys. Rev. D} {\bfseries 109} no.~11, (2024) 112003}, \href{http://arxiv.org/abs/2402.08668}{{\ttfamily arXiv:2402.08668 [hep-ex]}}.

\bibitem{Darme:2021gtt}
L.~Darm\'e, B.~Fuks, and F.~Maltoni, ``{Top-philic heavy resonances in four-top final states and their EFT interpretation},'' \href{http://dx.doi.org/10.1007/JHEP09(2021)143}{{\em JHEP} {\bfseries 09} (2021) 143}, \href{http://arxiv.org/abs/2104.09512}{{\ttfamily arXiv:2104.09512 [hep-ph]}}.

\bibitem{Miralles:2019uzg}
V.~Miralles and A.~Pich, ``{LHC bounds on colored scalars},'' \href{http://dx.doi.org/10.1103/PhysRevD.100.115042}{{\em Phys. Rev. D} {\bfseries 100} no.~11, (2019) 115042}, \href{http://arxiv.org/abs/1910.07947}{{\ttfamily arXiv:1910.07947 [hep-ph]}}.

\bibitem{ATLAS:2020zzb}
{\bfseries ATLAS} Collaboration, G.~Aad {\em et~al.}, ``{Search for dijet resonances in events with an isolated charged lepton using $\sqrt{s} = 13$ TeV proton-proton collision data collected by the ATLAS detector},'' \href{http://dx.doi.org/10.1007/JHEP06(2020)151}{{\em JHEP} {\bfseries 06} (2020) 151}, \href{http://arxiv.org/abs/2002.11325}{{\ttfamily arXiv:2002.11325 [hep-ex]}}.

\bibitem{CMS:2019gwf}
{\bfseries CMS} Collaboration, A.~M. Sirunyan {\em et~al.}, ``{Search for high mass dijet resonances with a new background prediction method in proton-proton collisions at $\sqrt{s} =$ 13 TeV},'' \href{http://dx.doi.org/10.1007/JHEP05(2020)033}{{\em JHEP} {\bfseries 05} (2020) 033}, \href{http://arxiv.org/abs/1911.03947}{{\ttfamily arXiv:1911.03947 [hep-ex]}}.

\bibitem{ATLAS:2024xdc}
{\bfseries ATLAS} Collaboration, G.~Aad {\em et~al.}, ``{Combination of searches for singly produced vector-like top quarks in pp collisions at $\sqrt{s} = 13$ TeV with the ATLAS detector},'' \href{http://arxiv.org/abs/2408.08789}{{\ttfamily arXiv:2408.08789 [hep-ex]}}.

\bibitem{ATLAS:2022tla}
{\bfseries ATLAS} Collaboration, G.~Aad {\em et~al.}, ``{Search for pair-produced vector-like top and bottom partners in events with large missing transverse momentum in pp collisions with the ATLAS detector},'' \href{http://dx.doi.org/10.1140/epjc/s10052-023-11790-7}{{\em Eur. Phys. J. C} {\bfseries 83} no.~8, (2023) 719}, \href{http://arxiv.org/abs/2212.05263}{{\ttfamily arXiv:2212.05263 [hep-ex]}}.

\bibitem{Mandal:2012rx}
T.~Mandal and S.~Mitra, ``{Probing Color Octet Electrons at the LHC},'' \href{http://dx.doi.org/10.1103/PhysRevD.87.095008}{{\em Phys. Rev. D} {\bfseries 87} no.~9, (2013) 095008}, \href{http://arxiv.org/abs/1211.6394}{{\ttfamily arXiv:1211.6394 [hep-ph]}}.

\bibitem{Mandal:2016csb}
T.~Mandal, S.~Mitra, and S.~Seth, ``{Probing Compositeness with the CMS $eejj$ and $eej$ Data},'' \href{http://dx.doi.org/10.1016/j.physletb.2016.05.020}{{\em Phys. Lett. B} {\bfseries 758} (2016) 219--225}, \href{http://arxiv.org/abs/1602.01273}{{\ttfamily arXiv:1602.01273 [hep-ph]}}.

\bibitem{Das:2015lna}
K.~Das, S.~Majhi, S.~K. Rai, and A.~Shivaji, ``{NLO QCD corrections to the resonant Vector Diquark production at the LHC},'' \href{http://dx.doi.org/10.1007/JHEP10(2015)122}{{\em JHEP} {\bfseries 10} (2015) 122}, \href{http://arxiv.org/abs/1505.07256}{{\ttfamily arXiv:1505.07256 [hep-ph]}}.

\bibitem{vonWeizsacker:1934nji}
C.~F. von Weizsacker, ``{Radiation emitted in collisions of very fast electrons},'' \href{http://dx.doi.org/10.1007/BF01333110}{{\em Z. Phys.} {\bfseries 88} (1934) 612--625}.

\bibitem{Williams:1934ad}
E.~J. Williams, ``{Nature of the high-energy particles of penetrating radiation and status of ionization and radiation formulae},'' \href{http://dx.doi.org/10.1103/PhysRev.45.729}{{\em Phys. Rev.} {\bfseries 45} (1934) 729--730}.

\bibitem{Budnev:1975poe}
V.~M. Budnev, I.~F. Ginzburg, G.~V. Meledin, and V.~G. Serbo, ``{The Two photon particle production mechanism. Physical problems. Applications. Equivalent photon approximation},'' \href{http://dx.doi.org/10.1016/0370-1573(75)90009-5}{{\em Phys. Rept.} {\bfseries 15} (1975) 181--281}.

\bibitem{Kane:1984bb}
G.~L. Kane, W.~W. Repko, and W.~B. Rolnick, ``{The Effective W+-, Z0 Approximation for High-Energy Collisions},'' \href{http://dx.doi.org/10.1016/0370-2693(84)90105-9}{{\em Phys. Lett. B} {\bfseries 148} (1984) 367--372}.

\bibitem{Dawson:1984gx}
S.~Dawson, ``{The Effective W Approximation},'' \href{http://dx.doi.org/10.1016/0550-3213(85)90038-0}{{\em Nucl. Phys. B} {\bfseries 249} (1985) 42--60}.

\bibitem{Chanowitz:1985hj}
M.~S. Chanowitz and M.~K. Gaillard, ``{The TeV Physics of Strongly Interacting W's and Z's},'' \href{http://dx.doi.org/10.1016/0550-3213(85)90580-2}{{\em Nucl. Phys. B} {\bfseries 261} (1985) 379--431}.

\bibitem{Gribov:1972ri}
V.~N. Gribov and L.~N. Lipatov, ``{Deep inelastic e p scattering in perturbation theory},'' {\em Sov. J. Nucl. Phys.} {\bfseries 15} (1972) 438--450.

\bibitem{Lipatov:1974qm}
L.~N. Lipatov, ``{The parton model and perturbation theory},'' {\em Yad. Fiz.} {\bfseries 20} (1974) 181--198.

\bibitem{Altarelli:1977zs}
G.~Altarelli and G.~Parisi, ``{Asymptotic Freedom in Parton Language},'' \href{http://dx.doi.org/10.1016/0550-3213(77)90384-4}{{\em Nucl. Phys. B} {\bfseries 126} (1977) 298--318}.

\bibitem{Dokshitzer:1977sg}
Y.~L. Dokshitzer, ``{Calculation of the Structure Functions for Deep Inelastic Scattering and e+ e- Annihilation by Perturbation Theory in Quantum Chromodynamics.},'' {\em Sov. Phys. JETP} {\bfseries 46} (1977) 641--653.

\bibitem{Alwall:2014hca}
J.~Alwall, R.~Frederix, S.~Frixione, V.~Hirschi, F.~Maltoni, O.~Mattelaer, H.~S. Shao, T.~Stelzer, P.~Torrielli, and M.~Zaro, ``{The automated computation of tree-level and next-to-leading order differential cross sections, and their matching to parton shower simulations},'' \href{http://dx.doi.org/10.1007/JHEP07(2014)079}{{\em JHEP} {\bfseries 07} (2014) 079}, \href{http://arxiv.org/abs/1405.0301}{{\ttfamily arXiv:1405.0301 [hep-ph]}}.

\bibitem{Frederix:2018nkq}
R.~Frederix, S.~Frixione, V.~Hirschi, D.~Pagani, H.~S. Shao, and M.~Zaro, ``{The automation of next-to-leading order electroweak calculations},'' \href{http://dx.doi.org/10.1007/JHEP11(2021)085}{{\em JHEP} {\bfseries 07} (2018) 185}, \href{http://arxiv.org/abs/1804.10017}{{\ttfamily arXiv:1804.10017 [hep-ph]}}. [Erratum: JHEP 11, 085 (2021)].

\bibitem{Kilian:2007gr}
W.~Kilian, T.~Ohl, and J.~Reuter, ``{WHIZARD: Simulating Multi-Particle Processes at LHC and ILC},'' \href{http://dx.doi.org/10.1140/epjc/s10052-011-1742-y}{{\em Eur. Phys. J. C} {\bfseries 71} (2011) 1742}, \href{http://arxiv.org/abs/0708.4233}{{\ttfamily arXiv:0708.4233 [hep-ph]}}.

\bibitem{Moretti:2001zz}
M.~Moretti, T.~Ohl, and J.~Reuter, ``{O'Mega: An Optimizing matrix element generator},'' \href{http://arxiv.org/abs/hep-ph/0102195}{{\ttfamily arXiv:hep-ph/0102195}}.

\bibitem{Christensen:2010wz}
N.~D. Christensen, C.~Duhr, B.~Fuks, J.~Reuter, and C.~Speckner, ``{Introducing an interface between WHIZARD and FeynRules},'' \href{http://dx.doi.org/10.1140/epjc/s10052-012-1990-5}{{\em Eur. Phys. J. C} {\bfseries 72} (2012) 1990}, \href{http://arxiv.org/abs/1010.3251}{{\ttfamily arXiv:1010.3251 [hep-ph]}}.

\bibitem{Buckley:2014ana}
A.~Buckley, J.~Ferrando, S.~Lloyd, K.~Nordstr\"om, B.~Page, M.~R\"ufenacht, M.~Sch\"onherr, and G.~Watt, ``{LHAPDF6: parton density access in the LHC precision era},'' \href{http://dx.doi.org/10.1140/epjc/s10052-015-3318-8}{{\em Eur. Phys. J. C} {\bfseries 75} (2015) 132}, \href{http://arxiv.org/abs/1412.7420}{{\ttfamily arXiv:1412.7420 [hep-ph]}}.

\bibitem{Alloul:2013bka}
A.~Alloul, N.~D. Christensen, C.~Degrande, C.~Duhr, and B.~Fuks, ``{FeynRules 2.0 - A complete toolbox for tree-level phenomenology},'' \href{http://dx.doi.org/10.1016/j.cpc.2014.04.012}{{\em Comput. Phys. Commun.} {\bfseries 185} (2014) 2250--2300}, \href{http://arxiv.org/abs/1310.1921}{{\ttfamily arXiv:1310.1921 [hep-ph]}}.

\bibitem{Degrande:2014vpa}
C.~Degrande, ``{Automatic evaluation of UV and R2 terms for beyond the Standard Model Lagrangians: a proof-of-principle},'' \href{http://dx.doi.org/10.1016/j.cpc.2015.08.015}{{\em Comput. Phys. Commun.} {\bfseries 197} (2015) 239--262}, \href{http://arxiv.org/abs/1406.3030}{{\ttfamily arXiv:1406.3030 [hep-ph]}}.

\bibitem{Degrande:2011ua}
C.~Degrande, C.~Duhr, B.~Fuks, D.~Grellscheid, O.~Mattelaer, and T.~Reiter, ``{UFO - The Universal FeynRules Output},'' \href{http://dx.doi.org/10.1016/j.cpc.2012.01.022}{{\em Comput. Phys. Commun.} {\bfseries 183} (2012) 1201--1214}, \href{http://arxiv.org/abs/1108.2040}{{\ttfamily arXiv:1108.2040 [hep-ph]}}.

\bibitem{Shtabovenko:2020gxv}
V.~Shtabovenko, R.~Mertig, and F.~Orellana, ``{FeynCalc 9.3: New features and improvements},'' \href{http://dx.doi.org/10.1016/j.cpc.2020.107478}{{\em Comput. Phys. Commun.} {\bfseries 256} (2020) 107478}, \href{http://arxiv.org/abs/2001.04407}{{\ttfamily arXiv:2001.04407 [hep-ph]}}.

\bibitem{Shtabovenko:2016sxi}
V.~Shtabovenko, R.~Mertig, and F.~Orellana, ``{New Developments in FeynCalc 9.0}'' \href{http://dx.doi.org/10.1016/j.cpc.2016.06.008}{{\em Comput. Phys. Commun.} {\bfseries 207} (2016) 432--444}, \href{http://arxiv.org/abs/1601.01167}{{\ttfamily arXiv:1601.01167 [hep-ph]}}.

\bibitem{Mertig:1990an}
R.~Mertig, M.~Bohm, and A.~Denner, ``{FEYN CALC: Computer algebraic calculation of Feynman amplitudes},'' \href{http://dx.doi.org/10.1016/0010-4655(91)90130-D}{{\em Comput. Phys. Commun.} {\bfseries 64} (1991) 345--359}.

\bibitem{Hahn:2000kx}
T.~Hahn, ``{Generating Feynman diagrams and amplitudes with FeynArts 3},'' \href{http://dx.doi.org/10.1016/S0010-4655(01)00290-9}{{\em Comput. Phys. Commun.} {\bfseries 140} (2001) 418--431}, \href{http://arxiv.org/abs/hep-ph/0012260}{{\ttfamily arXiv:hep-ph/0012260}}.

\bibitem{Han:2003wu}
T.~Han, H.~E. Logan, B.~McElrath, and L.-T. Wang, ``{Phenomenology of the little Higgs model},'' \href{http://dx.doi.org/10.1103/PhysRevD.67.095004}{{\em Phys. Rev. D} {\bfseries 67} (2003) 095004}, \href{http://arxiv.org/abs/hep-ph/0301040}{{\ttfamily arXiv:hep-ph/0301040}}.

\bibitem{ATLAS:2019nkf}
{\bfseries ATLAS} Collaboration, G.~Aad {\em et~al.}, ``{Combined measurements of Higgs boson production and decay using up to $80$ fb$^{-1}$ of proton-proton collision data at $\sqrt{s}=$ 13 TeV collected with the ATLAS experiment},'' \href{http://dx.doi.org/10.1103/PhysRevD.101.012002}{{\em Phys. Rev. D} {\bfseries 101} no.~1, (2020) 012002}, \href{http://arxiv.org/abs/1909.02845}{{\ttfamily arXiv:1909.02845 [hep-ex]}}.

\bibitem{CMS:2018gwt}
{\bfseries CMS} Collaboration, A.~M. Sirunyan {\em et~al.}, ``{Measurement and interpretation of differential cross sections for Higgs boson production at $\sqrt{s} =$ 13 TeV},'' \href{http://dx.doi.org/10.1016/j.physletb.2019.03.059}{{\em Phys. Lett. B} {\bfseries 792} (2019) 369--396}, \href{http://arxiv.org/abs/1812.06504}{{\ttfamily arXiv:1812.06504 [hep-ex]}}.

\bibitem{Buchkremer:2013bha}
M.~Buchkremer, G.~Cacciapaglia, A.~Deandrea, and L.~Panizzi, ``{Model Independent Framework for Searches of Top Partners},'' \href{http://dx.doi.org/10.1016/j.nuclphysb.2013.08.010}{{\em Nucl. Phys. B} {\bfseries 876} (2013) 376--417}, \href{http://arxiv.org/abs/1305.4172}{{\ttfamily arXiv:1305.4172 [hep-ph]}}.

\bibitem{Cacciapaglia:2012dd}
G.~Cacciapaglia, A.~Deandrea, L.~Panizzi, S.~Perries, and V.~Sordini, ``{Heavy Vector-like quark with charge 5/3 at the LHC},'' \href{http://dx.doi.org/10.1007/JHEP03(2013)004}{{\em JHEP} {\bfseries 03} (2013) 004}, \href{http://arxiv.org/abs/1211.4034}{{\ttfamily arXiv:1211.4034 [hep-ph]}}.

\bibitem{delAguila:2000rc}
F.~del Aguila, M.~Perez-Victoria, and J.~Santiago, ``{Observable contributions of new exotic quarks to quark mixing},'' \href{http://dx.doi.org/10.1088/1126-6708/2000/09/011}{{\em JHEP} {\bfseries 09} (2000) 011}, \href{http://arxiv.org/abs/hep-ph/0007316}{{\ttfamily arXiv:hep-ph/0007316}}.

\bibitem{Cacciapaglia:2010vn}
G.~Cacciapaglia, A.~Deandrea, D.~Harada, and Y.~Okada, ``{Bounds and Decays of New Heavy Vector-like Top Partners},'' \href{http://dx.doi.org/10.1007/JHEP11(2010)159}{{\em JHEP} {\bfseries 11} (2010) 159}, \href{http://arxiv.org/abs/1007.2933}{{\ttfamily arXiv:1007.2933 [hep-ph]}}.

\bibitem{Lee:1977eg}
B.~W. Lee, C.~Quigg, and H.~B. Thacker, ``{Weak Interactions at Very High-Energies: The Role of the Higgs Boson Mass},'' \href{http://dx.doi.org/10.1103/PhysRevD.16.1519}{{\em Phys. Rev. D} {\bfseries 16} (1977) 1519}.

\bibitem{MuCoL:2024oxj}
{\bfseries MuCoL} Collaboration, C.~Accettura {\em et~al.}, ``{MuCol Milestone Report No. 5: Preliminary Parameters},'' \href{http://arxiv.org/abs/2411.02966}{{\ttfamily arXiv:2411.02966 [physics.acc-ph]}}.

\bibitem{Guo:2023jkz}
Q.~Guo, L.~Gao, Y.~Mao, and Q.~Li, ``{Vector-like lepton searches at a muon collider in the context of the 4321 model*},'' \href{http://dx.doi.org/10.1088/1674-1137/ace5a7}{{\em Chin. Phys. C} {\bfseries 47} no.~10, (2023) 103106}, \href{http://arxiv.org/abs/2304.01885}{{\ttfamily arXiv:2304.01885 [hep-ph]}}.

\bibitem{Feldman:1997qc}
G.~J. Feldman and R.~D. Cousins, ``{A Unified approach to the classical statistical analysis of small signals},'' \href{http://dx.doi.org/10.1103/PhysRevD.57.3873}{{\em Phys. Rev. D} {\bfseries 57} (1998) 3873--3889}, \href{http://arxiv.org/abs/physics/9711021}{{\ttfamily arXiv:physics/9711021}}.

\bibitem{ILC:2013jhg}
{\bfseries ILC} Collaboration, ``{The International Linear Collider Technical Design Report - Volume 2: Physics},'' \href{http://arxiv.org/abs/1306.6352}{{\ttfamily arXiv:1306.6352 [hep-ph]}}.

\bibitem{CLIC:2018fvx}
{\bfseries CLIC} Collaboration, J.~de~Blas {\em et~al.}, ``{The CLIC Potential for New Physics},'' \href{http://arxiv.org/abs/1812.02093}{{\ttfamily arXiv:1812.02093 [hep-ph]}}.

\bibitem{Jungman:1995df}
G.~Jungman, M.~Kamionkowski, and K.~Griest, ``{Supersymmetric dark matter},'' \href{http://dx.doi.org/10.1016/0370-1573(95)00058-5}{{\em Phys. Rept.} {\bfseries 267} (1996) 195--373}, \href{http://arxiv.org/abs/hep-ph/9506380}{{\ttfamily arXiv:hep-ph/9506380}}.

\bibitem{Han:2020uak}
T.~Han, Z.~Liu, L.-T. Wang, and X.~Wang, ``{WIMPs at High Energy Muon Colliders},'' \href{http://dx.doi.org/10.1103/PhysRevD.103.075004}{{\em Phys. Rev. D} {\bfseries 103} no.~7, (2021) 075004}, \href{http://arxiv.org/abs/2009.11287}{{\ttfamily arXiv:2009.11287 [hep-ph]}}.

\bibitem{Han:2019grb}
T.~Han, A.~Ismail, and B.~Shams Es~Haghi, ``{SUSY Signals from QCD Production at the Upgraded LHC},'' \href{http://dx.doi.org/10.1016/j.physletb.2019.05.004}{{\em Phys. Lett. B} {\bfseries 793} (2019) 354--359}, \href{http://arxiv.org/abs/1902.05109}{{\ttfamily arXiv:1902.05109 [hep-ph]}}.

\bibitem{Han:2009ya}
T.~Han, I.~Lewis, and T.~McElmurry, ``{QCD Corrections to Scalar Diquark Production at Hadron Colliders},'' \href{http://dx.doi.org/10.1007/JHEP01(2010)123}{{\em JHEP} {\bfseries 01} (2010) 123}, \href{http://arxiv.org/abs/0909.2666}{{\ttfamily arXiv:0909.2666 [hep-ph]}}.

\bibitem{Pati:1974yy}
J.~C. Pati and A.~Salam, ``{Lepton Number as the Fourth Color},'' \href{http://dx.doi.org/10.1103/PhysRevD.10.275}{{\em Phys. Rev. D} {\bfseries 10} (1974) 275--289}. [Erratum: Phys.Rev.D 11, 703--703 (1975)].

\bibitem{Dobado:1987pj}
A.~Dobado, M.~J. Herrero, and C.~Munoz, ``{Production of Leptoquarks From Superstring Models in $e p$ Colliders},'' \href{http://dx.doi.org/10.1016/0370-2693(87)90638-1}{{\em Phys. Lett. B} {\bfseries 191} (1987) 449--455}.

\bibitem{Fritzsch:1981zh}
H.~Fritzsch and G.~Mandelbaum, ``{Weak Interactions as Manifestations of the Substructure of Leptons and Quarks},'' \href{http://dx.doi.org/10.1016/0370-2693(81)90626-2}{{\em Phys. Lett. B} {\bfseries 102} (1981) 319--322}.

\bibitem{Harari:1982xy}
H.~Harari, ``{Composite Models for Quarks and Leptons},'' \href{http://dx.doi.org/10.1016/0370-1573(84)90207-2}{{\em Phys. Rept.} {\bfseries 104} (1984) 159}.

\bibitem{Campana}
M.~Campana, {\em {Search for leptoquarks coupling to muons in lepton-quark collisions at LHC}}.
\newblock PhD thesis, Sapienza University of Rome, 2024.
\newblock \url{https://www.roma1.infn.it/exp/cms/tesiPHD/tesi_phd_completate/campana.pdf}.

\bibitem{Ruckl:1997ex}
R.~Ruckl, R.~Settles, and H.~Spiesberger, ``{Leptoquark pair production at $e^{+} e^{-}$ linear colliders: Signals and background},'' in {\em {Joint ECFA / DESY Study: Physics and Detectors for a Linear CollidfileTo be followed by 2nd workshop in July whose location and dates are to be determined, and 3rd workshop at DESY 20-22 Nov 1996)}}.
\newblock 9, 1997.
\newblock \href{http://arxiv.org/abs/hep-ph/9709315}{{\ttfamily arXiv:hep-ph/9709315}}.

\bibitem{Goncalves-Netto:2013nla}
D.~Goncalves-Netto, D.~Lopez-Val, K.~Mawatari, I.~Wigmore, and T.~Plehn, ``{Looking for leptogluons},'' \href{http://dx.doi.org/10.1103/PhysRevD.87.094023}{{\em Phys. Rev. D} {\bfseries 87} (2013) 094023}, \href{http://arxiv.org/abs/1303.0845}{{\ttfamily arXiv:1303.0845 [hep-ph]}}.

\bibitem{Almeida:2022udp}
E.~d.~S. Almeida, A.~Alves, O.~J.~P. \'Eboli, and F.~S. Queiroz, ``{Resonant lepton-gluon collisions at the Large Hadron Collider},'' \href{http://dx.doi.org/10.1103/PhysRevD.107.055024}{{\em Phys. Rev. D} {\bfseries 107} no.~5, (2023) 055024}, \href{http://arxiv.org/abs/2212.06178}{{\ttfamily arXiv:2212.06178 [hep-ph]}}.

\bibitem{Hill:2002ap}
C.~T. Hill and E.~H. Simmons, ``{Strong Dynamics and Electroweak Symmetry Breaking},'' \href{http://dx.doi.org/10.1016/S0370-1573(03)00140-6}{{\em Phys. Rept.} {\bfseries 381} (2003) 235--402}, \href{http://arxiv.org/abs/hep-ph/0203079}{{\ttfamily arXiv:hep-ph/0203079}}. [Erratum: Phys.Rept. 390, 553--554 (2004)].

\bibitem{Chivukula:1995dt}
R.~S. Chivukula, R.~Rosenfeld, E.~H. Simmons, and J.~Terning, {\em {Strongly coupled electroweak symmetry breaking: Implication of models}}, \href{http://dx.doi.org/10.1142/9789812830265_0006}{pp.~352--382}.
\newblock 2, 1995.
\newblock \href{http://arxiv.org/abs/hep-ph/9503202}{{\ttfamily arXiv:hep-ph/9503202}}.

\bibitem{Dobrescu:2007xf}
B.~A. Dobrescu, K.~Kong, and R.~Mahbubani, ``{Leptons and Photons at the LHC: Cascades through Spinless Adjoints},'' \href{http://dx.doi.org/10.1088/1126-6708/2007/07/006}{{\em JHEP} {\bfseries 07} (2007) 006}, \href{http://arxiv.org/abs/hep-ph/0703231}{{\ttfamily arXiv:hep-ph/0703231}}.

\bibitem{Dobrescu:2007yp}
B.~A. Dobrescu, K.~Kong, and R.~Mahbubani, ``{Massive color-octet bosons and pairs of resonances at hadron colliders},'' \href{http://dx.doi.org/10.1016/j.physletb.2008.10.048}{{\em Phys. Lett. B} {\bfseries 670} (2008) 119--123}, \href{http://arxiv.org/abs/0709.2378}{{\ttfamily arXiv:0709.2378 [hep-ph]}}.

\bibitem{Frampton:1987dn}
P.~H. Frampton and S.~L. Glashow, ``{Chiral Color: An Alternative to the Standard Model},'' \href{http://dx.doi.org/10.1016/0370-2693(87)90859-8}{{\em Phys. Lett. B} {\bfseries 190} (1987) 157--161}.

\bibitem{Bagger:1987fz}
J.~Bagger, C.~Schmidt, and S.~King, ``{Axigluon Production in Hadronic Collisions},'' \href{http://dx.doi.org/10.1103/PhysRevD.37.1188}{{\em Phys. Rev. D} {\bfseries 37} (1988) 1188}.

\bibitem{Hill:1993hs}
C.~T. Hill and S.~J. Parke, ``{Top production: Sensitivity to new physics},'' \href{http://dx.doi.org/10.1103/PhysRevD.49.4454}{{\em Phys. Rev. D} {\bfseries 49} (1994) 4454--4462}, \href{http://arxiv.org/abs/hep-ph/9312324}{{\ttfamily arXiv:hep-ph/9312324}}.

\bibitem{Chivukula:1996yr}
R.~S. Chivukula, A.~G. Cohen, and E.~H. Simmons, ``{New strong interactions at the Tevatron?},'' \href{http://dx.doi.org/10.1016/0370-2693(96)00464-9}{{\em Phys. Lett. B} {\bfseries 380} (1996) 92--98}, \href{http://arxiv.org/abs/hep-ph/9603311}{{\ttfamily arXiv:hep-ph/9603311}}.

\bibitem{Dicus:2000hm}
D.~A. Dicus, C.~D. McMullen, and S.~Nandi, ``{Collider implications of Kaluza-Klein excitations of the gluons},'' \href{http://dx.doi.org/10.1103/PhysRevD.65.076007}{{\em Phys. Rev. D} {\bfseries 65} (2002) 076007}, \href{http://arxiv.org/abs/hep-ph/0012259}{{\ttfamily arXiv:hep-ph/0012259}}.

\bibitem{Idilbi:2009cc}
A.~Idilbi, C.~Kim, and T.~Mehen, ``{Factorization and resummation for single color-octet scalar production at the LHC},'' \href{http://dx.doi.org/10.1103/PhysRevD.79.114016}{{\em Phys. Rev. D} {\bfseries 79} (2009) 114016}, \href{http://arxiv.org/abs/0903.3668}{{\ttfamily arXiv:0903.3668 [hep-ph]}}.

\bibitem{Cao:2021qqt}
Q.-H. Cao, J.-N. Fu, Y.~Liu, X.-H. Wang, and R.~Zhang, ``{Probing top-philic new physics via four-top-quark production},'' \href{http://dx.doi.org/10.1088/1674-1137/ac0c6f}{{\em Chin. Phys. C} {\bfseries 45} no.~9, (2021) 093107}, \href{http://arxiv.org/abs/2105.03372}{{\ttfamily arXiv:2105.03372 [hep-ph]}}.

\bibitem{Han:2010rf}
T.~Han, I.~Lewis, and Z.~Liu, ``{Colored Resonant Signals at the LHC: Largest Rate and Simplest Topology},'' \href{http://dx.doi.org/10.1007/JHEP12(2010)085}{{\em JHEP} {\bfseries 12} (2010) 085}, \href{http://arxiv.org/abs/1010.4309}{{\ttfamily arXiv:1010.4309 [hep-ph]}}.

\bibitem{Kaplan:2008ie}
D.~E. Kaplan, K.~Rehermann, M.~D. Schwartz, and B.~Tweedie, ``{Top Tagging: A Method for Identifying Boosted Hadronically Decaying Top Quarks},'' \href{http://dx.doi.org/10.1103/PhysRevLett.101.142001}{{\em Phys. Rev. Lett.} {\bfseries 101} (2008) 142001}, \href{http://arxiv.org/abs/0806.0848}{{\ttfamily arXiv:0806.0848 [hep-ph]}}.

\bibitem{Chivukula:2013xla}
R.~S. Chivukula, A.~Farzinnia, J.~Ren, and E.~H. Simmons, ``{Hadron Collider Production of Massive Color-Octet Vector Bosons at Next-to-Leading Order},'' \href{http://dx.doi.org/10.1103/PhysRevD.87.094011}{{\em Phys. Rev. D} {\bfseries 87} no.~9, (2013) 094011}, \href{http://arxiv.org/abs/1303.1120}{{\ttfamily arXiv:1303.1120 [hep-ph]}}.

\bibitem{Kileng:1994kw}
B.~Kileng and P.~Osland, ``{Gluino production in electron - positron annihilation},'' \href{http://dx.doi.org/10.1007/BF01556378}{{\em Z. Phys. C} {\bfseries 66} (1995) 503--512}, \href{http://arxiv.org/abs/hep-ph/9407290}{{\ttfamily arXiv:hep-ph/9407290}}.

\bibitem{Kileng:1994vc}
B.~Kileng and P.~Osland, ``{Light gluino production at LEP},'' in {\em {9th International Workshop on High-energy Physics and Quantum Field Theory}}, pp.~44--50.
\newblock 9, 1994.
\newblock \href{http://arxiv.org/abs/hep-ph/9411248}{{\ttfamily arXiv:hep-ph/9411248}}.

\bibitem{Berge:2003bz}
S.~Berge, \href{http://dx.doi.org/10.3204/DESY-THESIS-2003-048}{``{Gluino and squark pair production at future linear colliders},''} master thesis, Hamburg U., 12, 2003.

\end{thebibliography}\endgroup
\end{document}